\documentclass[11pt,eqsecnum]{article}

\usepackage{amsmath, amssymb}
 \usepackage{epsfig}
\usepackage{mathrsfs}
\usepackage{psfrag}
\usepackage{graphicx}
\usepackage{epstopdf}

\usepackage{color}

\usepackage{color}

\newcommand \be{\begin{eqnarray}}
\newcommand \ee{\end{eqnarray}}

\newcommand{\ba}{\begin{eqnarray}}
\newcommand{\ea}{\end{eqnarray}}

\def\l {\lambda}

\def \td {\tilde}
\def \td {\tilde}
\def\sn{{\rm sn}}
\def\cn{{\rm cn}}
\def\dn{{\rm dn}}

\def \F {{_F}}
 \def\F{\mathbb{F}}

\def\p{\phi}
\def\b{\beta}
\def \g{\gamma}

\def\S{{\cal S}}
\def\r{\rho}

\def\k{\kappa}
\def\E{{\cal E}}

\def\s{\sigma}

 \def\tg{\tilde{\g}}

 \def\J{{\cal J}}
\def \adss  {$AdS_5 \times S^5$}
\def \sql {\sqrt{\lambda}}

\begin{document}
\renewcommand{\thefootnote}{\arabic{footnote}}

\def \foot {\footnote}
\def \bi{\bibitem}

\def \tr {{\rm tr}}
\def \ha {\frac{1}{2}}

\def \ci{\cite}
\def \N {{\mathcal N}}
\def \const {{\rm const}}
\def \t {\tau}
\def\S{{\mathcal S} }
\def \nn{\nonumber\\}
\def \XX {{\rm X}}
\def \fl {\sqrt[4]{\l}}

 \def \vp {\varphi} \def \bs {\bar \s }
\def \k {\kappa}
\def\foot{\footnote}
\def \ci {\cite}
\def \ov {\over}

\def \bp {\begin{pmatrix}}  \def \epm {\end{pmatrix}}
\def \ha {{\textstyle{1 \ov 2}}}

\def \bi {\bibitem}
\newcommand{\lar}{\longrightarrow}
\def \la {\label}
\def \Tr  {{\rm Tr}}

\def \T {{\cal T}}
\def \l {\lambda}
\def\foot{\footnote}
\def \tl  {{\tilde \l}}
\def \sql {{\sqrt \l}}
\def \adss {$AdS_5 \times S^5$\ }
\newcommand{\rf}[1]{(\ref{#1})}

\def \bp {\begin{pmatrix}}
 \def \emp {\end{pmatrix}}

 \def \dett  {{\det}}

\def \qr {{\hat \rho}}
\def \const {{\rm const}}
\def \bea{\begin{eqnarray}}
\def \eea{\end{eqnarray}}
\def \no {\nonumber}
\def \tr {{\rm Tr}}
\def \g {\gamma}
\def \tm {\mbb{T}}
\def \ha {\fr{ 1}{ 2}}
\def \half {\fr{ \trm{1}}{\trm{2}}}
\def \s {\sigma}
\def \vp {\varphi}
\def \td {\tilde}
\def \z {\zeta}
\def \H {{\rm H}}
\def \Tr {{\rm Tr}}
\def \ep {\epsilon}
\def \bp {\begin{pmatrix}}
 \def \emp {\end{pmatrix}}
\def \ef {\end{document}}
\def \del {\partial}
\def \G {\Gamma} \def \ha { { 1 \ov 2}}  \def \tg  {\td \Gamma} \def \m {\mu}
 \def \tdb {\bar }
\def \lm {Lam\'e\ }
 \def \tdb {\bar }
 \def\Z{\mathbb{Z}}
 \def\K{\mathbb{K}}
\def \ed {\end{document}}

\def \bi{ \bibitem}

 \newcommand{\ads}{AdS_5\times S^5}
 \def \sn {{\rm sn}} \def \bK {{\mathbb{K}}}

\let\5=\overline

\overfullrule=0pt
\parskip=2pt
\parindent=12pt
\headheight=0in \headsep=0in \topmargin=0in \oddsidemargin=0in

\vspace{ -3cm} \thispagestyle{empty} \vspace{-1cm}
 \vspace{-1cm}
\begin{flushright} Imperial-TP-AT-2010-03
\end{flushright}
\begin{center}
{\Large\bf Exact computation of one-loop correction to energy \\
\vspace{0.1cm}
of pulsating strings in $AdS_5 \times S^5$ }

 \vspace{0.8cm} {
  M.~Beccaria$^{a,}$\footnote{matteo.beccaria@le.infn.it},
  G.~V.~Dunne$^{b,}$\footnote{dunne@phys.uconn.edu},
  G.~Macorini$^{a,}$\footnote{guido.macorini@le.infn.it},
  A.~Tirziu$^{c,}$\footnote{atirziu@aps.org} and
  A.~A.~Tseytlin$^{d,}$\footnote{Also at
 Lebedev  Institute, Moscow.\ \
  tseytlin@imperial.ac.uk
 }}\\
 \vskip  0.5cm

\small
{\em
$^{a}$
Physics Department, Salento University and INFN, 73100 Lecce, Italy

 \vskip 0.05cm
$^{b}$ Department of Physics,  University of Connecticut,
Storrs CT 06269-3046, USA

\vskip 0.05cm
$^{c}$  American Physical Society, 1 Research Road,
Ridge, NY 11961, USA

\vskip 0.05cm
$^{d}$  The Blackett Laboratory, Imperial College,
London SW7 2AZ, U.K. }

\normalsize
\end{center}

 \vskip 0.8cm

 \begin{abstract}
In the  present  paper, which is  a  sequel to arXiv:1001:4018,
 we  compute the  one-loop correction to the energy of   pulsating
string solutions in $AdS_5\times S^5$.
We show that, as for rigid spinning string elliptic
solutions, the fluctuation
  operators for pulsating   solutions can be
also put into the single-gap  Lam\'e form.
A novel aspect of  pulsating solutions
is that the one-loop  correction to their energy
is expressed   in  terms of the stability angles
of the quadratic fluctuation operators.
We explicitly study the  ``short string''
 limit of the corresponding one-loop  energies,
 demonstrating a certain universality of the
 form  of the energy of ``small'' semiclassical strings.
 Our results  may help  to shed light on the structure
 of strong-coupling expansion
  of anomalous dimensions of dual gauge  theory operators.

\end{abstract}

\newpage

\tableofcontents

\renewcommand{\theequation}{1.\arabic{equation}}
 \setcounter{equation}{0}
\setcounter{footnote}{0}
\setcounter{section}{0}

\def \N {{\cal N}}\def \E {{\cal E}}
\def \PP  {{\mathscr{P}}}

\section{Introduction}

\adss string energies (or planar $\cal N$=4 SYM anomalous dimensions)
 are,  in general,  complicated functions of string
tension (or `t Hooft coupling) and various
charges. They should   be described by the
integrability-based Thermodynamic Bethe Ansatz (see, e.g.,
\ci{reviews} for a review)
 but detailed patterns of their behaviour  with coupling and charges
 are still  poorly
understood.

The semiclassical string  expansion  applies in
 a particular  limit when charges  scale as string
 tension $\frac{\sql}{2 \pi} \gg 1$.
It was argued   \cite{grom} that in this limit  the TBA
prediction matches exactly the 1-loop string correction to the energies.
However, the details of the correspondence between the  Bethe
ansatz (or algebraic curve)
 description of the  semiclassical solutions and their
direct 2d  string sigma model  description remain to be
  clarified  for various  non-trivial types of solutions.
 Also, following \cite{tt,rt}
one  may  hope to shed light on  the structure
of anomalous dimensions of ``short''  operators by
studying
the  ``short string'' limit --  the limit of small values of
the semiclassical
parameters.   Assuming that this limit commutes with the
 large $\sql$   limit  one could then  interpolate
the result to small (fixed)  values of the  quantum string
 charges.

Apart from the case of the
 rational rigid   string solutions for which the
fluctuation Lagrangian has
constant coefficients \ci{ft03},  the  direct 2d quantum field theory
 computation  of
the 1-loop correction to string
energies  is difficult. To compute the one-loop energy   one
 needs to find  the spectrum of  mixed-mode
 fluctuation operators which are
  second order matrix 2d
differential  operators with coordinate-dependent  coefficients.
As was  explained in the previous paper of three of us  \cite{bd},
for the next to the simplest case of  elliptic solutions  with
the  folded spinning
string  being the  basic example,
  one can  compute the corresponding determinants using the special
  Lam\'e
form of the    fluctuation operators.

The present  paper is a natural sequel to \cite{bd} where we treat
other  cases of  similar elliptic solutions -- pulsating
string solutions in $AdS_5$ and $S^5$.
A novel aspect of pulsating solutions
-- which are time-dependent  rather than rigid stationary as in
previous  spinning string case --
is that the 1-loop  correction to their energy
is determined  in a more complicated way  than just by summing
characteristic frequencies.
As we shall discuss below, in
  this case one needs to follow the general semiclassical method
  of quantization of time-periodic solitons   \ci{dhn,viced}
expressing the  correction to the energy in terms of the stability angles
of the fluctuation operators.

The structure of this  paper is as  follows.
In section 2  we
discuss a simple  pulsating string solution on $S^2$.
We present the bosonic and fermionic quadratic fluctuation operators and show that these
can be written as 1-d  differential operators
with single-gap Lam\'e potentials.

In section 3 we  repeat the same for   a     pulsating string in $AdS_3$.
In section 4  we review  the general semiclassical quantization
approach focussing on the case  of
 time-dependent potentials,  where one needs to  use the  stability angles
 to compute one-lopp correction to the energy.

In section 5  we
use the  data of the previous sections
to write down  the complete
 one-loop correction to the energy of pulsating string solutions.
We then  derive  the explicit expansion of the energy
in the
 short string limit.

In section   6 we return to a rigid spinning string case of a type
considered in \cite{bd} --
 the folded string in $\mathbb{R} \times S^2$.
 We find  the exact one-loop correction
to its energy   and expand it explicitly
 in the short string limit.

Finally, in section 7 we
 present our conclusions and
 compare  the short-string  one-loop  energies
for all the four cases  of the  folded and pulsating strings in $AdS_5 \times S^5$
studied in \cite{bd}  and here.

 Our notations and some
details of the  computation of fluctuation operators
are presented in appendices A--D.
 Appendix E supplements the discussion in section 4
 presenting  a heuristic derivation
 of the one-loop expression for the  energy in terms
 of the stability angles.
 In Appendix F  we discuss an alternative  antiperiodic choice for the
  fermionic boundary
 condition mentioned in section 7.


\renewcommand{\theequation}{2.\arabic{equation}}
 \setcounter{equation}{0}
\setcounter{section}{1}

\section{Pulsating string  in $\mathbb{R}\times S^{2}$}

In this section we shall  start  with the classical   string background
representing pulsating string  in $\mathbb{R}\times S^{2}$
and  then consider the second order 2d  operators governing the spectrum of small fluctuations near this solution.
We shall then demonstrate that as in the case of another elliptic solution -- folded string in
$AdS_3$  considered  in \ci{bd}
 these operators can be put  in the  Lam\'e form, and thus their spectrum and
 eventually the one-loop correction to the string
 energy   can be computed exactly.

\subsection{Classical solution}

The pulsating string solution in ${\mathbb R} \times S^2$,  which is a generalization of a circular pulsating string on a
plane   was considered in \ci{gkp} and
  in \cite{m03} starting  with the Nambu action in the static gauge.
   Here we review this solution in the conformal gauge following \ci{krut}.
    Let us start with the following ansatz for the bosonic string  coordinates in
     $\mathbb{R} \times S^2$  ($m=1,2,...$)
\be t = \kappa\,\tau, \qquad \psi=\psi(\tau), \qquad \phi=m\,\sigma\ , \qquad
ds^{2} = -dt^{2}+d\psi^{2}+\sin^{2}\psi\,d\phi^{2}  \ .\la{yllq}
\ee
The equation of motion and the   conformal gauge constraint (which
 implies the former for $\dot\psi\not=0$)
are
\be \ddot\psi+m^{2}\,\sin\psi\cos\psi = 0 \ , \ \ \ \ \ \
\dot\psi^{2}+m^{2}\,\sin^{2}\psi = \kappa^{2} \ .\la{llq}
\ee
The solution with $\psi(0)=0$ can be written in terms of the Jacobi elliptic function \cite{ww}
\be\la{soll}
\sin\psi(\tau) = \frac{\kappa}{m}\,{\rm sn}\Big(m\,\tau\,|\,\frac{\kappa^{2}}{m^{2}}\Big),
\qquad |\sin\psi| \le \sin\psi_{0} = \frac{\kappa}{m}.
\ee
To  have a time-periodic solution we need to assume  $\kappa < m$.
The induced metric and its curvature are
\begin{equation}
ds^2= m^2 \sin^2 \psi\, \eta_{ab} \ , \ \ \ \ \ \ \ R^{(2)}= \frac{2}{m^2 \sin^2 \psi}(m^2 \sin^2 \psi- \frac{\kappa^2}{\sin^2 \psi}).
\end{equation}

The energy and the oscillation  number $N= \frac{\sqrt{\lambda}}{2\pi}\oint d\psi \ \dot \psi $
(the adiabatic invariant associated to $\psi$) are\foot{$\frac{\sql}{2
\pi}$ is string tension. We follow the same notation for elliptic functions as in \ci{bd}.}
\ba
\mathcal E_0 &=& \frac{E}{\sqrt\lambda} = \kappa\ ,\la{pqo} \\
\mathcal N &=& \frac{N}{\sqrt\lambda} = \int^{2\pi}_0 \frac{d\psi}{2\pi} \sqrt{\kappa^{2}-m^{2}\,\sin^{2}\psi} =
\int_{0}^{4\mathbb{K}(\frac{\kappa^{2}}{m^{2}})}\frac{d\tau}{2\pi} \Big(\kappa^{2}-m^{2}\,\sin^{2}\psi\Big).
\la{wpqo}\ea
A short calculation gives
\be\la{nan}
\mathcal N =  \frac{2m}{\pi}\Big[\Big(\frac{\kappa^{2}}{m^{2}}-1\Big)\,\mathbb{K}\Big(\frac{\kappa^{2}}{m^{2}}\Big)
+\mathbb{E}\Big(\frac{\kappa^{2}}{m^{2}}\Big)\Big]\ .
\ee
where ${\mathbb K}$ and ${\mathbb E}$ are the usual elliptic functions \cite{ww,bd}.
The condition $\kappa < m$ gives an upper bound for $\mathcal{N}$, i.e.  here,
like for the  folded string in ${\mathbb R} \times S^2$
(but in   contrast to the folded
string in $AdS_3$)  one cannot take the large $\mathcal{N}$ limit.

 The expansion of $\mathcal N$  for  small $\kappa$ gives
\be
\mathcal N = \frac{\kappa ^2}{2 m}+\frac{\kappa ^4}{16 m^3}+\frac{3 \kappa ^6}{128 m^5}+\dots \ .
\la{nen}
\ee
Thus the  short string or small oscillation number (${\mathcal N} \to 0$)
 expansion of  the classical energy   is
\be
\mathcal E_0(\mathcal N) = \sqrt{2\,m\,\mathcal N}\,\Big(
1-\frac{\mathcal N}{8 m}-\frac{5 \mathcal N^2}{128 m^2}+\dots
\Big). \la{ses}
\ee

\subsection{Quadratic   fluctuation Lagrangian }

In order to compute the $1$-loop correction to energy \rf{ses}
we need to find the  operators of quadratic
fluctuations. The derivation is standard  with  details  presented in Appendix B
(we follow Appendix A of \cite{mtt06}).


The bosonic  fluctuation operators   can be found directly in conformal gauge
where  we find
 two mixed modes. They can be
decoupled  by solving the Virasoro constraints  with the resulting  fluctuation action
(with 2 ``longitudinal'' massless  modes omitted) being equivalent to the
one that can be found directly using  the static gauge as in \ci{bd}.

The conformal gauge fluctuations in $AdS_5$ directions  are
represented by  a free massless  ``ghost''  field plus four free  massive fields
with  mass $ \kappa$
(here $k=1,2,3,4$;\ \  $\del_a \del^a = - \del_\tau^2 + \del_\s^2$)
\be\la{kop}
L_{AdS}^{(2)} =- \frac{1}{2}( \dot \b^2 - \b'^2) + \frac{1}{2}
( \dot y_k^2 - y'_k{}^2   -  {\kappa^2}  y_k y_k
 ) \ .
\ee
The Lagrangian for the  five  $S^5$ fluctuations ($ \xi,\eta,  z_1,z_2,z_3 $)
 is
\ba
&& L_{S}^{(2)} =
  \frac{1}{2}(\dot \xi^2-\xi'^2-M_\xi^2\,\xi^2) + \frac{1}{2}(\dot \eta^2-\eta'^2-M_\eta^2\,\eta^2)
 +\ m\,\cos\psi\,( \xi\,\eta'-\xi'\,\eta) \cr
 && \ \ \ \ \ \ \ \ \ \ \ \ \ \ +\
\frac{1}{2}(  \dot z_i^2  -   z'_i{}^2     - M^2 z_i^2    ) \ , \la{laga}
\ea
where the background-dependent masses  are
\ba
M^2    =\kappa^{2}-2m^{2}\,\sin^{2}\psi, \qquad
M_\xi ^2    = \kappa^{2}+m^{2}\cos(2\psi) \ , \qquad M_\eta^2 =m^{2}\,\cos(2\psi)   \ .  \la{mas}
\ea
Solving the Virasoro constraints  one  can show that the  coupled system  $(\xi,\eta$)
is equivalent  to a decoupled system of one  massless mode and
of the  massive mode with the Lagrangian
\ba\la{ggg}
L=
\frac{1}{2}(\dot g^2-g'^2-\td M^2 \, g^2) \  , \ \ \ \ \ \ \ \ \ \ \ \
\td M ^2= \kappa^{2}\big(1-\frac{2}{\sin^{2}\psi}\big)   \ . \ea
This is the same  fluctuation  Lagrangian (with the  massless  modes omitted)
as found by starting with the Nambu action and
imposing the static gauge  on the fluctuations (see Appendix A).
An equivalent fluctuation action follows  also  from  the Pohlmeyer
reduction approach \cite{Iwashita:2010tg}.


The fermionic fluctuation Lagrangian is found as, e.g., in \ci{ft02,bd}.
In the standard $\theta^1 = \theta^2$ kappa symmetry gauge  it is
(cf. Appendix A)
\be\la{fef}
{\cal L}_F = -2\,i\,\overline\vartheta\Big(-\rho^a\,D_a-\frac{i}{2}\varepsilon^{ab}\,
\rho_a\,\Gamma_*\,\rho_b\Big)\,\vartheta \ ,
\ee
 leading  to the following expression for the fermionic fluctuation operator (see Appendix B)
\be\la{ffe}
D_{F} = \Gamma_{0}\partial_{\tau}-\Gamma_{9}\,\partial_{\sigma}+\Gamma_{079} \dot\psi \ .
\ee
Since  we are interested in its eigenvalues and
 determinant, we can take the square of the simpler operator
\be\la{fufe}
\tilde D_F \equiv  \Gamma_{09}\,D_{F} = \Gamma_{9}\partial_{\tau}-\Gamma_{0}\,\partial_{\sigma}
-\Gamma_{7} \dot\psi \ .
\ee
Diagonalizing $\Gamma_{97}$ (i.e. replacing it by $\pm i$)  we  get the
following second order fermionic operator
\be\la{sqf}
\tilde D_F^{2}{}_\pm  = \partial_{\tau}^{2}-\partial_{\sigma}^{2}+ M^2_\pm \ , \ \ \ \ \ \ \ \ \ \ \ \
M^2_\pm= \dot\psi^{2}\pm i\,\ddot\psi.
\ee
A simple check on the  resulting fluctuation Lagrangian
is provided by demonstrating the  UV
finiteness of the
1-loop  partition function.
In conformal gauge that requires  showing that the  sum of the effective
 mass-squared terms for bosons equals that for the fermions.\foot{The contribution of the
two  mixed
fluctuations  can be found   by  rewriting  the corresponding  terms as
$
A'^{2}+B'^{2}+\mu\,AB'
= (A'-\frac{\mu}{4}B)^{2}+(B'+\frac{\mu}{4}A)^{2}-\frac{\mu^{2}}{16}(A^{2}+B^{2})$
and observing that the  ``connection'' terms do not produce  UV divergences.}
We  find  that the sum of the physical  4+4  bosonic and 4+4 fermionic  effective   mass squared terms\foot{For this counting argument we may ignore
the $\pm i\,\ddot\psi$ terms in the fermionic masses in \rf{sqf},
as they sum up to zero.}
 \ba
AdS &:& 4\times \kappa^{2}, \nn
S^{5} &:& 3\times (\kappa^{2}-2m^{2}\sin^{2}\psi), \nn
&& 1\times (m^{2}\cos(2\psi)-m^{2}\cos^{2}\psi), \nn
&& 1\times (\kappa^{2}+m^{2}\cos(2\psi)-m^{2}\cos^{2}\psi), \nn
F &:& -8\times(\kappa^{2}-m^{2}\sin^{2}\psi)
\ea
 indeed sums to  zero.
 In the static gauge we get
\ba
AdS &:& 4\times \kappa^{2}, \nn
S^{5} &:& 3\times (\kappa^{2}-2m^{2}\sin^{2}\psi), \nn
&& 1\times \kappa^{2}\,(1-\frac{2}{\sin^{2}\psi}), \nn
F &:& -8\times(\kappa^{2}-m^{2}\sin^{2}\psi) \la{rwrr}
\ea
and the  sum is
\be
2m^{2}\sin^{2}\psi-2\frac{\kappa^{2}}{\sin^{2}\psi} = \sqrt{-g}\,R^{(2)}  \ , \la{rwr}
\ee
where $\sqrt{-g}\,R^{(2)}$  is proportional to the   Euler density of  the induced metric as
expected  on general grounds \ci{dgt}.
 As discussed in \ci{bd},  integrated   over the 2-space,
 this is proportional to the Euler number which vanishes for the
  cylinder topology under discussion.



%

\subsection{Remarks on single-gap Lam\'e operator}

We will show in the next section that each of the
above quadratic fluctuation operators can be transformed into the
``single-gap Lam\'e" form
\be\label{lamegen2}
\Big[-\partial^2_x+\  2k^2\,{\rm sn}^2(x\,|\,k^2)\Big]\,f(x)=\Lambda\,f(x) \quad ,
\ee
for suitable choices of the
coordinate $x$ and the
elliptic parameter $k^2$. This fact is significant because equation (\ref{lamegen2}) has simple solutions and properties, which we review briefly here (see also \cite{ww,bd}).
The two independent Bloch solutions of (\ref{lamegen2}) are
\be\label{solslame}
f_{\pm}(x) = \frac{H(x\pm\alpha)}{\Theta(x)}\,e^{\mp\,x\, Z(\alpha)} \ ,
\ee
where $H, \Theta, Z$ are the Jacobi Eta, Theta and Zeta functions \cite{ww},
 and the spectral parameter $\alpha=\alpha(\Lambda)$ is related to the
 eigenvalue $\Lambda$ by the transcendental equation:
\be\label{alphaeq}
{\rm sn}(\alpha\,|\,k^2) = \sqrt{\frac{1+k^2-\Lambda}{k^2}}\ .
\ee
Using the periodicity properties of the Jacobi functions we see that the Bloch solutions $f_\pm(x)$ acquire a phase under a
shift through one period $2{\mathbb K}$:
\be\label{solperiodic}
f_{\pm}(x+2\mathbb{K})=-f_{\pm}(x)\,e^{\mp\,2\,\mathbb{K}\, Z(\alpha)}\,\equiv\,f_{\pm}(x)\,e^{2i\,
\mathbb{K}\,p(\alpha)}\ .
\ee
This defines the quasi-momentum as
\be\label{momentums}
p(\Lambda) = i\, Z(\alpha\,|\,k^2)+\frac{\pi}{2\,\mathbb{K}} \ .
\ee
As explained in \cite{bd}, knowing
an explicit expression for the quasi-momentum implies that we
 can write an explicit expression for the
  corresponding  determinant of the fluctuation operator. We
  will return to this in sections  4 and 5.




\subsection{Lam\'e  form of  fluctuation operators}

{Having motivated the significance of the single-gap Lam\'e form
 of the fluctuation operators, we will now present their explicit form for each
of the decoupled (static gauge)
 fluctuation operators
in  the  pulsating string case.
Since the fluctuation potentials are independent of $\sigma$ for the
pulsating string solutions, we may use  the  Fourier decomposition of the
 $\sigma$ dependence,
$X(\tau, \sigma) = X(\tau)\ e^{in\sigma}$, so that $- \del_\tau^2 +
 \del_\s^2  +  M^2(\tau)\to  - \del_\tau^2 +    M^2(\tau) - n^2 $.
Depending on the form of the mass term (i.e. potential) $M^2(\tau)$,
we  find three types of Lam\' e operators , which we  discuss
in turn.  }

\

{\bf Type I operator}

{The operator associated to the three $S^{5}$ modes $z_i$ in \rf{laga}
 with mass $M^{2} = \kappa^{2}-2m^{2}\sin^{2}\psi$
is
\be\la{qqpa}
{\cal O}_{I} = -\partial_{\tau}^{2}+2m^{2}\sin^{2}\psi-\kappa^{2}-n^{2}\ .
\ee
Taking into account the specific form of the solution $\psi(\tau)$ in \rf{soll}, it  can be written as
\ba
&&{\cal O}_{I} = m^{2}\,\Big[-\partial^2_x+\  2k^2\,{\rm sn}^2(x\,|\,k^2)-\Lambda\Big], \la{o1} \\
&&
x = m\,\tau, \qquad {k^2 = \frac{\kappa^2}{m^2}}, \qquad \Lambda =
 \frac{\kappa^{2}+n^{2}}{m^{2}}  \  ,  \la{xxx}
\ea
{which is of  the single-gap Lam\'e form in (\ref{lamegen2}).}
The classical stability region is the set of $\Lambda$ for which the quasi-momentum is real
and the solution of the differential equation is quasi-periodic. Outside this region, the solution is unbounded
and is not acceptable. The Lam\'e operator in (\ref{o1}) is called ``single-gap" because there are just two allowed bands separated by a single gap:
\be
\Lambda\in[k^{2},1]\cup[k^{2}+1,+\infty] \ .
\ee
Assuming $\kappa<m$ this  gives
\be
n\in[0, \sqrt{m^{2}-\kappa^{2}}]\cup[m,+\infty] \ .
\ee
Notice that for $\kappa \le \sqrt{2m-1}$ the above range covers all
 integers $n$. If this condition is not satisfied,
there are certain values of $n$ which give rise to unstable
fluctuations.}\foot{Let us mention that
a hybrid string solution with two spins in $AdS_5$ and
pulsating in $S^5$ was considered before in \cite{kl05}.
The numerical analysis in \cite{kl05}
showed that generally  pulsation improves
the stability of a spinning string.}

\

 {\bf Type II operator}

{Next, consider the $S^{5}$ mode in \rf{ggg}
 with mass  $\td M^{2} = \kappa^{2}\big(1-\frac{2}{\sin^{2}\psi}\big)$, i.e.
with the associated operator
 \be \la{oqpq}
 {\cal O}_{II} = -\partial_{\tau}^{2}+\frac{2\kappa^{2}}{\sin^2\psi}-\kappa^{2}-n^{2}\ .
 \ee
 After using \rf{soll}  and definitions in \rf{xxx} we get
\be \la{opa}
{\cal O}_{II} = m^{2}\,\Big[-\partial^2_x+\  2\,{\rm ns}^2(x\,|\,k^2)-\Lambda\Big]  \ .
\ee
Taking into account  the identity,
$
{\rm ns}(z\,|\,k^{2}) = k\,{\rm sn}(z+i\,\mathbb{K}'\,|\,k^{2})
$,
we can write\footnote{{We use the standard notation
${\mathbb K}^\prime(k^2)\equiv {\mathbb K}(1-k^2)$.}}
\ba\la{topa}
{\cal O}_{II} &=& m^{2}\,\Big[-\partial^2_{x}+\  2\,k^{2}\,{\rm sn}^2(x\,|\,k^2)-\Lambda\Big],\\
x &\equiv&  m\,\tau+i\,\mathbb{K}^\prime \quad, \quad k=\frac{\kappa}{m}\quad, \quad \Lambda = \frac{\kappa^{2}+n^{2}}{m^{2}}\quad ,
\ea
which is again of the single-gap Lam\'e form {in (\ref{lamegen2}).}}

\

{\bf Type III operator}


The fermion fluctuation operator in  \rf{sqf}
 with the  mass {$M^2_\pm = \dot\psi^{2}\pm i\, \ddot\psi$}
leads to
 \be
 {\cal O}_{III}^\pm = -\partial_{\tau}^{2}-\dot\psi^{2}\mp i\,\ddot \psi-n^{2}.
 \ee
 Using  the explict  form of $\psi(\tau)$ in \rf{soll} and \rf{xxx} we get
\ba
{\cal O}_{III}^\pm &=&
m^{2}\,\Big[-\partial^2_x-k^{2}\,{\rm cn}^2(x\,|\,k^2)
\mp i\, k\,{\rm sn}(x\,|\,k^2)\,{\rm dn}(x\,|\,k^2)-\frac{n^{2}}{m^{2}}\Big] \ ,
\la{jjj} \\
x &\equiv&  m\,\tau \ , \quad k=\frac{\kappa}{m}\ , \quad \Lambda = \frac{n^{2}}{m^{2}}\ .
\ea
This operator is non-hermitian, but  is PT-symmetric \cite{bender} and has a real spectrum.\footnote{Notice
 also that after a shift of $x$ by
$i\,\mathbb{K}'$ the potential
is real and singular. This is another way to show that its
spectrum is real.}
Moreover, while it does not look like the standard single-gap Lam\'e operator, it can be
transformed into this form by a combination of {rescaling of $x$ and a Gauss
 transformation of the elliptic parameter $k^2$,}
\ba
&& {\cal O}_{III}^\pm =
\bar{m}_\pm^{2}\,\Big[-\partial^2_x+2\,\bar{k}_\pm^{2}\,{\rm sn}^2(\bar{x}\,|\,\bar{k}_\pm^2)
-\Lambda\Big] \ ,  \la{kol}
\\
&& \bar x \equiv \bar{m}_\pm\,\tau+\frac{1}{2} {{\mathbb K}(\bar{k}^2_\pm)} \ ,
\qquad \bar m_\pm =\frac{m}{2}\Big(\sqrt{1-\frac{\kappa^2}{m^2}}\pm i\frac{\kappa}{m}\Big)\  , \la{pyy}\\
&&
\bar{k}_\pm^2=\pm 4\frac{ \frac{i\,\kappa}{m} \sqrt{1-\frac{\kappa^2}{m^2}}}
{\Big( {\sqrt{1-\frac{\kappa^2}{m^2}}} \pm {\frac{i\kappa}{m}}\Big)^2}\ , \qquad
 \Lambda = \frac{n^{2}}{\bar{m}_\pm^{2}+\bar{k}_\pm^2}\ .
\la{jjj2}
\ea
Thus
 we  again  find a fluctuation operator
  of the single-gap Lam\'e form {in (\ref{lamegen2})}.

\renewcommand{\theequation}{3.\arabic{equation}}
 \setcounter{equation}{0}
\setcounter{section}{2}

\section{Pulsating string  in $AdS_{3}$}

Our  aim here will be  to repeat the discussion of the previous section in the case of
the
pulsating string solution in $AdS_{3}$ discussed in \cite{m03,ptt05}.

\subsection{Classical solution}

Using the standard parametrization of  the $AdS_{5}$ metric
\be\la{b}
ds^2_{AdS_5} = d\rho^2-\cosh^2\rho\, dt^2 + \sinh^2\rho\, (d\theta^2+\cos^2\theta\,
 d\phi_1^2 + \sin^2\theta \, d\phi_2^2)\ ,
\ee
let us  look for a string  solution in conformal gauge assuming
\be\la{a}
t=t(\tau), \qquad \rho=\rho(\tau), \qquad \theta=0, \qquad \phi_{1}=m\sigma, \qquad \phi_{2} = 0\ .
\ee
The non-trivial conformal gauge constraint  and the two equations of motion are
\ba
&&\dot\rho^{2}-\cosh^{2}\rho\,\dot t^{2}+m^{2}\,\sinh^{2}\rho=0\ , \\
&&
2\sinh\rho\,\dot\rho\,\dot t+\cosh\rho\,\ddot t =0\ ,      \qquad
\ddot\rho+\sinh\rho\cosh\rho\,(m^{2}+\dot t^{2}) =0  \ . \la{eqw}
\ea
The first equation in \rf{eqw}  can be integrated and expressed in terms of
 the 
  integral of motion ${\cal E}_0$ (global AdS energy)
\be
\dot t = \frac{{\cal E}_0}{\cosh^{2}\rho}\ .  \la{glo}
\ee
Then the  conformal gauge constraint becomes
\be
\dot\rho^{2}-\frac{{{\cal E}_0}^{2}}{\cosh^{2}\rho}+m^{2}\,\sinh^{2}\rho=0\ ,
\ee
with its derivative implying  the second-order equation for $\rho$.
Its   solution with $\rho(0)=0$
 is\footnote{The equation for $x = \sinh\rho$ is
$\dot x^{2} = m^{2}(x^{2}-R_{-})(R_{+}-x^{2})$ and can be compared with the differential equation for
the elliptic function
${\rm sd}(z|m)$.}
\ba
&&\sinh\rho(\tau) = \sqrt{\frac{-R_{+}R_{-}}{R_{+}-R_{-}}}\ {\rm sd}\Big(m\sqrt{R_{+}-R_{-}}\,\tau
\,|\, \frac{R_{+}}{R_{+}-R_{-}}\Big)\ ,\la{kww}
\\
&&
R_{\pm} = \frac{-m\pm\sqrt{m^{2}+4\,\mathcal E_0^{2}}}{2m}\ . \la{ras}
\ea
An alternative  form of the solution is
\ba\la{aal}
&&\sinh \rho(\tau)= \sqrt{R_{+}}\,{\rm cn}\big(x+\bK(k^2)\,|\,k^2\big)
\ , \\
&&x=m \sqrt{R_{+}-R_{-}}\ \tau\ \equiv w\, \tau \  , \la{xerx} \\
&&  k^2=\frac{R_{+}}{R_{+}-R_{-}} {=\frac{1}{2}\Big(1-\frac{1}{\sqrt{1+
\left(\frac{2 \mathcal{E}_0}{m}\right)^2}}\Big) \  . }
\la{yty} \ea
The induced metric and its curvature are found to be
\begin{equation}
ds^2= m^2 \sinh^2 \rho\  \eta_{ab}, \qquad \quad R^{(2)}=-2 - \frac{2 \mathcal{E}_0^2}{m^2 \sinh^4 \rho}\
.
\end{equation}
The oscillation  number is defined as follows
\begin{equation}
N=\sqrt{\lambda}\ \N \ , \ \ \  \ \ \   \N  = \frac{1}{2\pi}\oint d\rho\
\dot {\rho}=\frac{2\sqrt{\lambda}}{\pi}\int_{0}^{\rho_{max}}d\rho
\sqrt{\frac{\mathcal{E}_0^2}{\cosh^2 \rho}-m^2 \sinh^2 \rho} \ . \la{ner}
\end{equation}
Changing the  variable to $x=\sinh \rho$ we get
\ba
\N
&=&\frac{2m \sqrt{\lambda}}{\pi}\int_0^{\sqrt{R_{+}}}\frac{d
x}{1+x^2}\sqrt{(R_{+}-x^2)(x^2-R_{-})}\la{net}\\
&=&
\frac{2m \sqrt{\lambda}}{\pi}\frac{1}{\sqrt{-R_{-}}}\bigg[R_{-} \mathbb{E}(q) +(1+R_{+})\big[\mathbb{K}(q)
-(1+R_{-}) \Pi(-R_{+},q)\big]\bigg], \quad \quad q=\frac{R_{+}}{R_{-}} \ . \nonumber
\end{eqnarray}
In the short string limit
when  $\E_0$ and $\N$  are small  we find (cf. \rf{ses})
\footnote{
For large $\N$ we get Eq.~(3.19) of  \cite{m03} (see also Appendix C  of  \cite{ptt05}). At small $\N$
one should get (C.14) of  \cite{ptt05},
\  $E = \sqrt{4\sqrt\lambda N}+\dots$. These relations show that the definition of $\N$
in  \cite{ptt05} as well as  in \cite{m03} is off by a factor of 2. Indeed, in the
 flat space case one should have
$E = \sqrt{2\sqrt\lambda N}+\dots$, where $T=  \frac{\sqrt\lambda}{2 \pi}$ is the tension.
 This $N$ is identified with the total oscillator number which in the closed string case
has to be even. For instance,  in the
bosonic string  case $\alpha' p^{2} = M^{2} = 2(N_{L}+N_{R}-2) = 2(N-1)$ with
$N=N_{L}+N_{R}$.}
\ba
\mathcal N &=&
\frac{\mathcal E_0^2}{2 m}-\frac{5 \mathcal E_0^4}{16 m^3}+\frac{63 \mathcal E_0^6}{128 m^5}
-\frac{2145 \mathcal E_0^{8}}{2048
   m^7}+\dots \ , \la{nenn} \\
\mathcal E_0 &=& \sqrt{2m\,\mathcal N}\,\left(
1+\frac{5 \mathcal N}{8 m}-\frac{77  \mathcal N^2}{128 m^2}+\frac{1365  \mathcal N^3}{1024 m^3}+\dots
\right)\ .\la{eeee}
\ea

\subsection{Quadratic fluctuation Lagrangian}

Using the  conformal gauge we get  5  massless modes in $S^5$
and  the following bosonic quadratic fluctuation Lagrangian for $AdS_5$ modes
 \begin{eqnarray}
\tilde{L}&=&\frac{1}{2} \bigg[\sinh^2 \rho\ [ (\partial_a \tilde{\beta}_1)^2
-m^2 \tilde{\beta}_1^2]+
 \sinh^2 \rho \cos^2 m \sigma \ (\partial_a \tilde{\beta}_3)^2
\nonumber\\
&+& \sinh^2 \rho \ (\partial_a \tilde{\beta}_2)^2
- \cosh^2 \rho\  (\partial_a \tilde{t})^2
 + 4 \mathcal{E}_0 \tanh \rho \ \tilde{\rho} \partial_\tau  \tilde{t} \nonumber\\
&+& (\partial_a \tilde{\rho})^2+(m^2+\frac{\mathcal{E}_0^2}{\cosh^4 \rho}) \cosh (2 \rho)\ \tilde{\rho}^2+ 2 m \sinh
(2 \rho)\ \tilde{\rho} \ \partial_\s \tilde{\beta}_2\bigg]
\end{eqnarray}
After the field redefinitions (here $\r= \r(\tau)$, cf. \rf{a})
\begin{equation}
 \cos m \s \  \sinh \rho\ \tilde{\beta}_3 = \eta_1, \quad \quad  \sinh \rho\ \tilde{\beta}_1
 = \eta_2, \quad \quad \cosh \rho\
 \tilde{t}= \zeta, \quad \quad \sinh \rho  \ \tilde{\beta_2} = \chi \ ,
\end{equation}
the fluctuation Lagrangian becomes ($i=1,2$)
\begin{eqnarray}
&&\tilde{L}=\frac{1}{2}\bigg[
  (\partial_a \eta_i)^2
 + 2 m^2 \sinh^2 \rho\  \eta^2_i
+ (\partial_a \chi)^2 + m^2 (2 \sinh^2 \rho+1)\chi^2 +4 m \cosh \rho \ \tilde{\rho} \ \del_\s\chi \nonumber\\
&& + \ (\partial_a \tilde{\rho})^2 +(m^2+\frac{\mathcal{E}_0^2}{\cosh^4 \rho}) \cosh (2 \rho)\ \tilde{\rho}^2
-(\partial_a \zeta)^2   +  (\frac{\mathcal{E}_0^2}{\cosh^4 \rho}-2 m^2 \sinh^2 \rho)\  \zeta^2  \nonumber\\
&& - \ 4 \mathcal{E}_0 \frac{\sinh \rho}
{\cosh^2 \rho}\ \zeta \partial_\tau  \tilde{\rho}-4 \mathcal{E}_0
\frac{\dot{\rho}}{\cosh^3 \rho}\ \zeta \tilde{\rho} \  \bigg] \ .
\end{eqnarray}
Like in the folded string case in $AdS_3$
 the fluctuation $\tilde{\rho}$ couples
to  two other fluctuations.
  As in the pulsating or folded string cases,  to  decouple the fluctuations one 
 needs
to use  the Virasoro constraints expanded at first order in the fluctuations
or to use the  static gauge on the fluctuations.
In the latter case we get 5 massless  and 2+1   massive modes with
the  following Lagrangian   (see Appendix C for details)
\begin{eqnarray}\la{stt}
\tilde{L}=\frac{1}{2}\bigg[
 (\partial_a \eta_i)^2+ 2 m^2  \sinh^2 \rho\ \eta_i^2
   +  (\partial_a \psi)^2+  \big(2 m^2 \sinh^2 \rho - \frac{2 \mathcal{E}_0^2}{\sinh^2 \rho}\big)\ \psi^2 \bigg]
\ .    \end{eqnarray}
The fermionic  fluctuation operator that follows from \rf{fef} is given by (cf. \rf{ffe}; see Appendix C)
\be\la{jf}
D_{F}' = \Gamma_{0}\partial_{\tau}-\Gamma_{3}\partial_{\sigma}+m\,\Gamma_{124}\, \sinh\rho\, \
.
\ee
Squaring it  and diagonalizing we get (cf. \rf{sqf})
\be \la{squf}
(D_{F}')^2  \to     \mathcal{O}_{F}=
 -\partial_\tau ^2 + \partial_\s^2 -m^2 \sinh^2 \rho \pm i m  \cosh  \rho\  \dot \rho
\ .
\ee
As for the $\mathbb{R}\times S^{2}$ pulsating string we can then check UV finiteness
 either by
 computing  the sum of the  squares of effective  masses in conformal
gauge  (absorbing the mixing terms in the covariant derivatives),
 or  by computing  the sum of the squares of  masses in
the static
gauge,  checking that the result is proportional to the Euler number density.
In the conformal gauge we find  that the non-zero   (mass)$^2$ terms are
\ba
\eta &:& 2m^{2}\sinh^{2}\rho\no \\
\xi &:& 2m^{2}\sinh^{2}\rho\no \\
\chi &:& m^{2}(2\sinh^{2}\rho+1)-\frac{1}{16}(4m\cosh\rho)^{2}\no \\
\rho &:& (m^{2}+\frac{{\cal E}_0^{2}}{\cosh^{2}\rho}) \cosh(2\rho)-\frac{1}{16}(4{\cal E}_0
\frac{\sinh\rho}{\cosh^{2}\rho})^{2}-\frac{1}{16}(4m\cosh\rho)^{2}\no \\
\zeta &:& -(\frac{{\cal E}_0^{2}}{\cosh^{4}\rho}-2m^{2}\sinh^{2}\rho )-\frac{1}{16}(4{\cal E}_0
\frac{\sinh\rho}{\cosh^{2}\rho})^{2}\no\\
F &:& -8m^{2}\sinh^{2}\rho
\ea
which indeed  sum up to  zero.\foot{We used that to compute
 the trace of the square of the  mass  matrix  we may ignore the  contributions
from mixing terms without derivatives of the fluctuating fields.
Also, we
took into account that the time-like fluctuation
 $\zeta$ has the opposite (ghost)  sign of the kinetic term so that  $\zeta\to i\zeta$ is
required in order to  bring it to canonical normalization, i.e. we should  set
$m_{\zeta}^{2}\zeta^{2}\to -m_{\zeta}^{2}\zeta^{2}$.}
In the static gauge  the non-zero (mass)$^2$ terms are
\ba
2 &\times& 2m^{2}\sinh^{2}\rho \no\\
1 &\times& 2m^{2}\sinh^{2}\rho-\frac{{2{\cal E}_0^{2}}}{\sinh^{2}\rho}\no\\
-8 &\times& m^{2}\sinh^{2}\rho
\ea
and  their sum has the same   value  as in \rf{rwr}
\be
-2(m^{2}\sinh^{2}\rho+\frac{{\cal E}_0^{2}}{\sinh^{2}\rho}) = \sqrt{-g}\,R^{(2)} \ .\la{rwwr}
\ee
 Again, the $\tau$-integral  of this term  vanishes upon taking into account the boundary contributions
 (the world-sheet topology is that of a cylinder that has  zero Euler number).

\subsection{Lam\' e form of fluctuation operators}

Let us now show that as in the previous section the fluctuation operators in the static gauge \rf{stt},\rf{squf}
can be put into the standard Lam\'e form.


The bosonic operator  with mass
$M^2=2 m^2 \sinh^2 \rho$ in \rf{stt}
 can be put into the type I  Lam\'e form as follows ($\del_\s \to i n$)
{\begin{eqnarray}\la{ones}
&&\mathcal{O}_{I}= w^2 \big[-\partial_{x}^2 + 2k^2 \sn^2(x |k^2)-\Lambda \big] \ ,\\
&&x\equiv w\, \tau+ \mathbb{K}(k^2)\ , \qquad \Lambda=2k^2+\frac{n^2}{w^2}\ , \\
&&k^2=\frac{R_{+}}{R_{+}-R_{-}} =\frac{1}{2}\Big(1-\frac{1}{\sqrt{1+\big(\frac{2
 \mathcal{E}_0}{m}\big)^2}}\Big) \ . \la{oqw}
\end{eqnarray}}
The  operator corresponding to the bosonic fluctuation in \rf{stt}  with the  mass
$
M^2=2m^2 \sinh^2 \rho - \frac{2 \mathcal{E}_0^2}{\sinh^2 \rho} $
can be written as
{\ba
&&\mathcal{O}_{II}= w^2 \big[-\partial_x^2 +
2k^2 \sn^2(x+\bK |k^2)+2 k^2 \sn^2(x+i\bK'|k^2)-A\big] \ , \la{kkl}\\
&&x\equiv w \,\tau\ , \qquad  A\equiv 4 k^2 + \frac{n^2}{m^2}
\frac{k^2}{R_{+}} =4 k^2 + \frac{n^2}{w^2}\ , \\
&&k^2=\frac{R_{+}}{R_{+}-R_{-}} =\frac{1}{2}
\Big(1-\frac{1}{\sqrt{1+\left(\frac{2 \mathcal{E}_0}{m}\right)^2}}\Big)  \ . \ea}
If written in terms of the   Weierstrass function the operator $\mathcal{O}_{II}$ is of the
finite gap  Lam\'e  form \cite{gw}. Explicitly, by using again a combination of
a Landen transformation  and a Jacobi imaginary transformation
it can be put into  the standard single-gap  Lam\'e form (\ref{lamegen2})
{\ba
 && \mathcal{O}_{II}= \frac{w^2}{\sqrt{1-p^2}} \Big[-\partial_{\tilde{x}}^2 +
 2\,p^2\, \sn^2(\tilde{x}|p^2) + B\Big]\ ,\la{pit}\\
 && \tilde{x}\equiv \frac{w\,\tau}{(1-p^2)^{1/4}}+i \,\bK'(p^2)\ , \qquad B=(2-A) \sqrt{1-p^2}-2\ ,
 \la{prt} \\
 &&
p^2 = 4 \Big[-2 k^2(k^2-1)- k (2k^2-1) \sqrt{k^2-1}\Big]   \ . \la{ltl}
\ea
Finally, the  fermionic fluctuation operator in \rf{squf} can be written as
\begin{eqnarray}
&& \mathcal{O}_{III}=w^2\Big[-\partial^2_{x}
-k^2 \cn^2(x|k^2)\ \mp i k\ \sn(x|k^2)\ \dn(x|k^2) -\frac{n^2}{w^2}\Big]  \ ,\la{frrff}\\
&& x=w\,\tau+\bK(k^2)\quad , \quad k^2=\frac{1}{2}\Big(1-\frac{1}
{\sqrt{1+\left(\frac{2 \mathcal{E}_0}{m}\right)^2}}\Big)  \ . \la{frf}
\end{eqnarray}
Note that this  operator is precisely of the
same form as the fermionic operator \rf{jjj}
in the ${\mathbb R} \times {\mathbb S}^2$ case,  with
 $m$ replaced by  $w$. Thus it can also be transformed
 into a  single-gap Lam\'e operator.}



\renewcommand{\theequation}{4.\arabic{equation}}
 \setcounter{equation}{0}
\setcounter{section}{3}

\section{Semiclassical quantization of time-periodic solutions of
integrable systems}

As a preparation for   computation of  1-loop correction to the  energy
 of the pulsating strings
here we shall
 briefly review the semiclassical quantization of general (classically integrable)
 Hamiltonian systems. We shall
 focus,  in particular, on the time-periodic case (see
\cite{viced} for details and references).

\subsection{Bohr-Sommerfeld-Maslov quantization}

Let us consider a classical Hamiltonian system on a space  $X$
($\dim X = n$)
 with a  Hamiltonian
$
H: T^{*}X\to \mathbb R.
$
We shall assume that its  quantum version is
 a self-adjoint operator $\widehat H$ on
 such that for $\hbar\to 0$ it reduces to
$H$.






The classical integrability requires the existence of $n$
functions $F_{1}, \dots, F_{n}\in C(T^{*}X)$ such that:
(i) $ dF_{1}\wedge\cdots \wedge dF_{n}\neq 0, \ {\rm almost\  everywhere},$
(ii) $\{F_{i}, F_{j}\}=0,$  and  (iii)
$H = H(F_{1}, \dots, F_{n})$.
This implies that the level sets
define $n$-tori (Liouville tori) foliating $T^{*}X$ and
invariant under the Hamiltonian flow. This allows one  to define the action
variables $I_{i}$ parametrizing the
foil base and the angle variables $\varphi_{i}$, the  coordinates of the torus.

{\it Semiclassical} integrability requires the existence of quantum extensions
$\widehat F_{i}$ of $F_i$  ($\widehat F_{i}\stackrel{\hbar\to 0} {\lar} F_{i}$)
such that in addition to  the condition (i)   above  they satisfy
(ii') $ [\widehat F_{i}, \widehat F_{j}]=\mathcal O(\hbar^{3})$   and (iii') $
\widehat H = H(\widehat F_{1}, \dots, \widehat F_{n})+\mathcal O(\hbar^{2})$.
Note  that $\widehat H$  is well defined without ordering problems
because of the condition (ii').
In general, the  condition of  {\it quantum} integrability is stronger as it
requires $ [\widehat F_{i}, \widehat F_{j}]=0$.

Suppose we  want to solve the joint diagonalization   problem
\be
\widehat F_{i}\,\psi = f_{i}\,\psi+\mathcal O(\hbar^{2})\ .\la{jo}
\ee
Under some technical simplifying assumption, a WKB-like solution exists if and only if
the following  Bohr-Sommerfeld-Maslov (BSM) quantization condition is satisfied  \cite{Voros}
\be
\frac{1}{2\pi\hbar}\int_{\gamma_{i}} {p}\cdot d  q = N_{i}+\frac{\mu_{i}}{4}+\mathcal O(\hbar), \qquad
i=1, \dots, n \ ,
\ee
where the  integers $N_i$  thus define  the action variables.
Here $\{\gamma_{i}\}$ is  a basis of  cycles of a Liouville torus
and the Maslov indices $\mu_{i}$ take into
account the critical points of the cycles. They generalize the familiar
 1/2  shift in the standard WKB
quantum mechanics relation where $\mu=2$ is the number of  inversion points.

If the classical invariant torus has only $p<n$ non trivial cycles, then it can be shown
that a simple change in the BSM
quantization condition is required.
It takes into account the fluctuations transverse to the codimension $p$
invariant torus.
In this  case the  quantization condition becomes
\be
\frac{1}{2\pi\hbar}\int_{\gamma_{k}} {p}\cdot d q =N_{k}+\frac{\mu_{k}}{4}
+\sum_{\alpha=p+1}^{n}\big(n_{\alpha}+\frac{1}{2}\big)\,\frac{\nu_{\alpha}^{(k)}}{2\pi}
+\mathcal O(\hbar),\quad \begin{array}{l} \ \ \ \ \ k=1, \dots, p, \\ \ \ \ \ \ n_{\alpha}\ll N_{k}\end{array}
\la{hhh}\ee
The {\it stability angles}\footnote{See below for their precise definition.}
$\nu_{\alpha}^{(k)}$ are
found  by studying
 the stability of small fluctuations
around the invariant torus (the condition $n_{\alpha}\ll N_{k}$ is necessary
in order  to  be able to use
 the linearised analysis).

These general considerations can be  applied to semiclassical quantization of
finite $g$-gap solutions of string theory
\cite{viced}.
 One starts with a   classical energy as a function of the  action  variables
 and then simply shifts them   according to the  BSM
 quantization
conditions, i.e.\foot{For finite gap  solutions   the Maslov indices are all equal to 2.}
\ba
 &&E = E_{cl}\Big(N_{1}\hbar+\frac{\hbar}{2}+\hbar\sum_{\alpha=g+2}^{\infty}\big(n_{\alpha}+\frac{1}{2}
\big)\frac{\nu_{\alpha}^{(1)}}{2\pi}, \dots,
\nn
&&\qquad\qquad
N_{g+1}\hbar+\frac{\hbar}{2}+\hbar\sum_{\alpha=g+2}^{\infty}\big(n_{\alpha}+\frac{1}{2}
\big)\frac{\nu_{\alpha}^{(g+1)}}{2\pi}\Big)+\mathcal O(\hbar^{2})\ . \la{eqe}
\ea
In particular, for the ground state  ($n_\alpha=0$)
 of a  1-gap superstring time-dependent
solution
of period $\T$, we can write (here $\hbar=\frac{1}{\sqrt\lambda}$, $\mathcal N = \frac{N}{\sqrt\lambda},
\ \mathcal E = \frac{E}{\sqrt\lambda}$)
\be
\mathcal E = \mathcal E_{cl}(\mathcal N) + \frac{1}{2\sqrt\lambda}\frac{1}{\T}\sum_{\nu_{s}>0 }\nu_{s}+
\mathcal O(\frac{1}{(\sqrt\lambda)^{2}}) \ . \la{pou}
\ee
Here $\T$ is the period of the solution which is the inverse of $\frac{d E}{d N}$. \foot{We took into account that for the superstring the balance
 of the number of the bosonic and fermionic fluctuations implies the
cancellation of the 1/2 shifts. We considered a single
$N$ in (\ref{eqe}) and expanded in small stability angles.
Additional details can be found
after Theorem 10.3.1 in  the second reference of \cite{viced}.}


In an integrable system,
 the stability angles  may  be computed directly
 since we can solve exactly the problem of evolution of a small
perturbation.
 This is due to the existence of a non-linear superposition principle
 that allows one to add a ``small'' solution on top of a soliton background by Backlund
 transformations.
The same construction can be carried over by adding a small additional cut
to a  finite cut solution of the corresponding integral equations
implied by the Bethe equations
(or more generally,   by considering a genus $g+1$ algebraic
curve infinitesimally near its  genus $g$  degeneration point).

The details of such  constructions (see \ci{gv})  may be quite involved and it is of
interest  to see   how they   compare  with
the  more standard approach based on second-order differential operators
 of small fluctuations near a solitonic solution.
The classical  integrability of the original system should translate into the special
properties of the corresponding fluctuation operators  given that they appear  upon
linearization of the same classical equations (see also \ci{bd}).


\subsection{Relation to  the Dashen-Hasslacher-Neveu  quantization prescription}

Let us now demonstrate the relation  between  the above approach and
the Dashen-Hasslacher-Neveu (DHN) approach \ci{dhn,hass}.\foot{See
 also  \ci{dorr} for applications
in the present string-theory context.}
DHN claimed   that one  is to   impose the condition (cf. \rf{hhh})
\be
\oint p\cdot dq +\sum_{s}(n_{s}+\frac{1}{2})(\T\frac{d\nu_{s}}{d\T}-\nu_{s}) = 2\pi N \ . \la{kkk}
\ee
Here we set $\hbar=1$ and denoted the stability angles by  $\nu_{s}$
which depend on the period $\T$.
In this condition the second term in the l.h.s. has to be considered as a perturbation
(it represents the one-loop correction in our case).
In the classical approximation
the  first term  is a function of the energy $E$
\be
\oint p\cdot dq = I(E).
\ee
Then the  inversion of the relation $I(E) = 2\pi N$ determines
the classical dependence of the energy on the action variable $N$
\be\la{ouiy}
E_{cl}(N)=
I^{-1}(2\pi N) \ .
\ee
For example, in the case of the pulsating string in ${\mathbb R}\times S^2$, discussed in Section 2,
the classical period is $\T=\frac{4}{m} {\mathbb K}\left(\frac{\kappa^2}{m^2}\right)$, and the classical action evaluated on the classical solution  is
\begin{eqnarray}
 S(\T)=
 \sqrt{\lambda} \int_0^\T d\tau\, L
 =
 2\,m\,\sqrt{\lambda}\Big[2\,{\mathbb E}\Big(\frac{\kappa^2}{m^2}\Big)+\Big(\frac{\kappa^2}{m^2}-2\Big) {\mathbb K}\Big(\frac{\kappa^2}{m^2}\Big)\Big]\ .
 \label{dhn-action}
 \end{eqnarray}
 Now let us make the Legendre transform from $S$ to $I\equiv \oint p\cdot dq=S-\T\frac{dS}{d\T}$. Using (\ref{dhn-action}), a short computation leads to the simple result: $dS/d\T=(dS/d\kappa^2)/(d\T/d\kappa^2)=-\sqrt{\lambda}\,\kappa^2/2$. Therefore, we find
  \begin{eqnarray}
 I=
 4\,m\,\sqrt{\lambda}\Big[{\mathbb E}\Big(\frac{\kappa^2}{m^2}\Big)+\Big(\frac{\kappa^2}{m^2}-1\Big) {\mathbb K}\Big(\frac{\kappa^2}{m^2}\Big)\Big]
 =2\, \pi\, N \ ,
 \end{eqnarray}
where $N$ is the adiabatic invariant defined in (\ref{wpqo}).

According to  DHN  \ci{dhn}, the quantum correction to the classical relation (\ref{ouiy}) can be expressed as follows. For  any $N$ in \rf{kkk}  the associated  energy
$E_{q}$ of a quantum state is
\be
E_{q} = E(N)+\sum_{s}(n_{s}+\frac{1}{2}) \frac{d\nu_{s}}{d\T} \ .\la{fr}
\ee
where $E(N)$ is to be obtained from (\ref{kkk}) by inverting $N=N(E)$.\footnote{The
l.h.s. in (\ref{kkk}) depends on the parameter $E$. The
 difference between $E(N)$ and $E_{cl}$ is due to the second
 term  in the l.h.s. in (\ref{kkk}).}

We can then solve \rf{kkk} perturbatively, i.e. $E(N)=E_{cl}(N)+\delta E$,
so that
\be
\left. \frac{dI}{dE}\right|_{E_{cl}}\delta E+\sum_{s}(n_{s}+\frac{1}{2})(\T\frac{d\nu_{s}}{d\T}-\nu_{s}) =0.
\ee
Using this in \rf{fr}  we get, expanding   to first order,
\ba
E_{q} &=& E_{cl}(N)-\frac{1}{\T}\sum_{s}(n_{s}+\frac{1}{2})(\T\frac{d\nu_{s}}{d\T}-\nu_{s})+
\sum_{s}(n_{s}+\frac{1}{2}) \frac{d\nu_{s}}{d\T} \nonumber \\
&=& E_{cl}(N)+\frac{1}{\T}\sum_{s}(n_{s}+\frac{1}{2}) \nu_{s}\  .\la{jko}
\ea
For $n_s=0$ (i.e. for the ``ground state'')
this   is  the  same
as the above expression  \rf{pou} found in \cite{viced}
and references therein.
It
 has an advantage of
  not involving  derivatives of the stability angles over the period.
This is the expression we shall use below.
An alternative  heuristic  derivation of (\ref{pou}) using
quantum field theory methods is sketched  in Appendix~E.



\renewcommand{\theequation}{5.\arabic{equation}}
 \setcounter{equation}{0}
\setcounter{section}{4}

\section{One-loop correction to energy of pulsating string solutions}

As discussed in the previous section, to compute the one-loop correction
 to the energy of pulsating string solutions we
need to find  the stability angles  for the Lam\'e-type  fluctuation operators
given in sections 2 and 3.

In general, given the 1-d spectral problem with a periodic potential
\begin{equation}\label{rder}
 \big[-\del_x^2 +V(x) \big]\,f(x) =\Lambda\, f(x) \ , \ \ \ \ \ \ \ \ \ \ \
 V(x+\T)=V(x)  \ ,
\end{equation}
its two independent solutions  $f_\pm (x)=e^{\pm i \, p(\Lambda) \, x}\,
\chi_\pm (x), \ \  \chi_\pm (x+\T)= \chi_\pm (x )$   satisfy
 \begin{eqnarray}
 f_\pm(x+\T)=e^{\pm i\nu }\ f_\pm(x) \ , \ \ \ \ \ \  \nu = p \T  \ ,
 \label{ntum}
 \end{eqnarray}
where $\nu$ is the ``stability angle'' and $p$ is the ``quasi-momentum''
  (in general,  $p$  is a function of $\Lambda, \T$ and a functional of $V$).

As noted in Sections 2 and 3, the relevant fluctuation operators studied in this paper are all of the single-gap Lam\'e form. Thus, from the quasi-periodicity properties of the explicit elliptic function solutions discussed in Section 2.3, we find exact expressions for the associated stability angles. These relations involve an auxiliary spectral parameter $\alpha$, and  solving   for this
parameter in an explicit  way appears to be complicated  in general.
Here we  will concentrate on the   ``short string'' or ``near-flat-space''
(small oscillation/energy) expansions of the exact relations.

\subsection{Pulsating string in $\mathbb{R}\times S^{2}$}

Let us recall that the period of the problem in section 2.1 is $\T=\frac{4\K}{m}$.
The short string limit  is the small $\kappa$ limit, in which the semiclassical oscillation
parameter $\mathcal{N}$ in \rf{nan}   is small.
Below we shall
consider the positive of the  two possible  stability angles differing by sign
 (see \cite{hass}).
 We shall also  rescale the stability angles by a factor of  $2\pi$.

Once stability angles  are computed, we will combine them  according to
\rf{pou} to find the  correction to the 2d energy.
Since in the present case the $AdS$ time $t$ and 2d  time $\tau$ are related as in
\rf{yllq},
i.e. $t = \k \tau$, there will be similar proportionality of  the periods,
and   the space-time energy  and the 2d energy will be related by
\be \la{repa}
E_{\rm spacetime} = \frac{1}{\k} E_{2d}  \ . \ee

\subsubsection{Stability angles}

The $4$  massless $ AdS_5$ fluctuations in \rf{kop}  have   the obvious stability angle
\be
\nu_{_{AdS_{5}}}  = 4\K\,\sqrt{k^{2}+\frac{n^{2}}{m^{2}}} \ , \ \ \ \ \ \ \ \
k\equiv \frac{ \kappa}{m}   \ . \la{io}
\ee
Expanding in small $\kappa$, i.e. in small $k$,   we get
\ba
\nu_{_{AdS_{5}}} &=& \frac{2 \pi  n}{m}+k^2 \left(\frac{\pi  m}{n}+\frac{\pi  n}{2 m}\right)+\frac{\pi  k^4 \left(-8 m^4+8 m^2 n^2+9 n^4\right)}{32 m
   n^3}\nonumber		\\
   && +\ \frac{\pi  k^6 \left(16 m^6-8 m^4 n^2+18 m^2 n^4+25 n^6\right)}{128 m n^5}+\dots\ . \  \la{ioi}
\ea
The $S^{5}$ bosonic fluctuations (both Type I \rf{qqpa} and Type  II \rf{oqpq}) are associated with the
 standard Lam\'e equation and the stability angle is
\be\la{cha}
&&\nu_{_{S^{5}}} = \pm 4\,\K\,\left(i\,\Z(\alpha\,|\,k^{2})+\frac{\pi}{2\K}\right) \equiv
 \pm 4\,\K\,i\,\Z(\alpha\,|\,k^{2})\ ,
\\
&&{\rm sn}(\alpha\,|\,k^{2}) = \sqrt\frac{1+k^{2}-\Lambda}{k^{2}} =
\frac{1}{k}\sqrt{1-\frac{n^{2}}{m^{2}}}\ .
\ee
We shall fix the sign in \rf{cha}   by the condition $\nu>0$.

Let us define $a=\sqrt{1-\frac{n^{2}}{m^{2}}}$ and begin with the case $|n|<|m|$ which is $a\in(0,1)$.
In general (here $\mathbb{E}=\mathbb{E}(k^{2})$, etc.),
\be \la{ivi}
\Z({\rm sn}^{-1}(\frac{a}{k}\,|\,k^{2})\,|\,k^{2}) = \int_{0}^{a/k}dt\,\left(
\sqrt\frac{1-k^{2}t^{2}}{1-t^{2}}-\frac{\mathbb{E}}{\K}\frac{1}{\sqrt{1-k^{2}t^{2}}\sqrt{1-t^{2}}}
\right).
\ee
In the short string limit $k\to 0^{+}$ limit, and for $a\in(0,1)$ we can exploit $\Z({\rm sn}^{-1}(1\,|\,k^{2})\,|\,k^{2})=0$. Taking into account Mathematica conventions for the cuts, we find
\be \la{ii}
\Z({\rm sn}^{-1}(\frac{a}{k}\,|\,k^{2})\,|\,k^{2}) = i\,\int_{1}^{a/k}\frac{dt}{\sqrt{t^{2}-1}}\,\left(
\sqrt{1-k^{2}t^{2}}-\frac{\mathbb{E}}{\K}\frac{1}{\sqrt{1-k^{2}t^{2}}}
\right).
\ee
The two basic integrals are
\ba
\int_{1}^{a/k}\frac{dt}{\sqrt{t^{2}-1}}\,
\sqrt{1-k^{2}t^{2}} &=& i\,(\mathbb{E}\Big(\arcsin\frac{a}{k}\,|\,k^{2})-\mathbb{E}\Big) \ , \\
\int_{1}^{a/k}\frac{dt}{\sqrt{t^{2}-1}}\,\frac{1}{\sqrt{1-k^{2}t^{2}}}
&=& i\,\Big(\F(\arcsin\frac{a}{k}\,|\,k^{2})-\K\Big). \label{bmfs}
\ea
In order to expand at small $k$, we use the transformation
\be
\mathbb{E}(\arcsin\frac{a}{k}\,|\,k^{2}) = \frac{\mathbb{E}}{\K} \,\F(\arcsin\frac{a}{k}\,|\,k^{2}) +i\,\sqrt{1-a^{2}}
\,\sqrt{1-\frac{k^{2}}{a^{2}}}\left(\frac{\Pi(a^{2}\,|\,k^{2})}{\K}-1\right).
\ee
The final result is remarkably simple since all incomplete elliptic integrals simplify. It reads
\be
\Z({\rm sn}^{-1}(\frac{a}{k}\,|\,k^{2})\,|\,k^{2})  = i\,\sqrt{1-a^{2}}
\,\sqrt{1-\frac{k^{2}}{a^{2}}}\left(1-\frac{\Pi(a^{2}\,|\,k^{2})}{\K}\right).
\ee
This expression can be expanded at small $k$. The product with $\K$ turns out to be
\ba  &&
i\,\K\,\Z({\rm sn}^{-1}(\frac{a}{k}\,|\,k^{2})\,|\,k^{2})  =
-\frac{1}{2} \pi  (\sqrt{1-a^2}-1)-\frac{\pi}{8}   \sqrt{1-a^2} k^2\no \\
&& \ -\frac{\pi  \sqrt{1-a^2} (9
   a^2+4) k^4}{128 a^2}
   -\frac{\pi  \sqrt{1-a^2} (25 a^4+12 a^2+8) k^6}{512 a^4}+{\cal O}\left(k^8\right).
\la{pyybis}
\ea
Using that $a=\sqrt{1-\frac{n^{2}}{m^{2}}}$
we find (for $n>0$) after fixing the sign in \rf{cha}  and subtracting  constant
$2\pi$ term
\ba
\label{eq:bose}
\nu_{_{S^5}} = -4\,i\,\K\,\Z({\rm sn}^{-1}(\frac{a}{k}\,|\,k^{2})\,|\,k^{2})  &=&
\frac{2\pi n}{m}
+\frac{\pi  k^2 n}{2 m}+\frac{\pi  k^4 n (13 m^2-9 n^2)}{32 m (m-n)
   (m+n)}\nonumber\\
   && +\ \frac{\pi  k^6 n (45 m^4-62 m^2 n^2+25 n^4)}{128 m (m-n)^2 (m+n)^2}+\dots
   \ea
Some comments are in order: (i)
 the singularity at $n=m$
is only  an apparent one since it happens at $a=0$ where our derivation cannot be applied
(the  above expression  following from
\rf{pyybis} is just zero at that
point; this is not a problem since our $n$ is discrete);
(ii) one can compare the l.h.s.  of \rf{eq:bose}  with the result (\ref{bmfs})
 for  imaginary $a$, i.e. for
  $n>m$,  and the equation  still applies.

\

Let us  mention  a different method to
 perform the short-string expansion    leading to \rf{eq:bose}:
 instead of expanding the exact
stability angles we shall use
direct perturbation theory  in small $\kappa$.
 We  shall use this perturbative approach  below in the case of
pulsating string in $AdS_3$.
Let us see how
this works for the Type I operator from \rf{o1}
 (here we take $m=1$ for simplicity), i.e.
\be
\mathcal O = -\partial_{x}^{2}+2\kappa^{2}\,{\rm sn}^{2}(x\,|\,\kappa^{2})-\kappa^{2}-n^{2} \
 . \la{kt}
\ee
Let us introduce the variable $y$ related to $x$ by
\be
x = \frac{2\K(\kappa^{2})}{\pi}\,y  \ ,
\ee
in terms of which the period of the problem
 is $\kappa$-independent   and equal to  $2\pi$.
 Expanding in $\k\to 0$ the  operator \rf{kt} can be written as
\ba
&&\mathcal O = \mathcal O_{0}+\kappa^{2}\mathcal O_{1}+\kappa^{4}\mathcal O_{2}+\cdots\ ,\\
&&\mathcal O_{0} =-\partial^{2}_{y}-n^{2}, \qquad
\mathcal O_{1} = \frac{1}{2}\partial^{2}_{y}-\cos 2y, \qquad
\mathcal O_{2} =\frac{3}{32}\partial^{2}_{y}+\sin^{2}y\cos^{2}y \ .
\ea
then one  easily finds iterative solution of
 $\mathcal O\ f=0$ in series expansion  in $\kappa^{2}$
starting from the  $f_0=e^{iny}$  solution of $\mathcal O_{0} f_0 =0$
\ba
&&f(y) = e^{iny+\kappa^{2}h_{1}(y)+\kappa^{4}h_{2}(y)+...} \ , \la{kwq} \\
&&
h_{1}(y) = \frac{i n \sin (2 y)}{4 (n^2-1)}-\frac{\cos (2 y)}{4 (n^2-1)}+\frac{i n y}{4}, \\
&&h_{2}(y) = \frac{1}{64 (n^2-1)^2} \Big[
 9 i n^5 y-22 i n^3 y+i n (n^2+1) \sin (4 y)+4 i (n^2-3) n \sin (2 y)\no \\
 && \ \ \ \ \ \ \ \ \ \ \ \ \ \ \ \ \ \ \ \ \ \  \ \ \ \ \ \ \ \ \ \ \ \ \ \ \ -2 n^2 \cos (4 y)+13 i n y+8 \cos (2
   y)\Big]\ .
\ea
The corresponding  stability angle is then
\be
\nu = -i\log\frac{f(2\pi)}{f(0)} = 2  \pi   n+\frac{1}{2}  \pi  \kappa^2 n+
\frac{ \pi  \kappa^4 n (9 n^2-13)}{32 (n^2-1)}+... \ ,
\ee
which  agrees with the expansion of the exact result (\ref{eq:bose})
after setting there $m=1$  and noting that $k=\frac{ \k}{m}$.

\


In the case of the
 fermionic fluctuation operator \rf{jjj}  the  expression for
  the stability angle is
\begin{equation}
\nu_{_F} = \pm 4\,i\,\mathbb{K}\,\Big[
\frac{1}{2}\,Z(\alpha({\beta})\,|\,k^{2})+
i\,\sqrt{\beta}\,\sqrt{1+\frac{16\,\beta\,k^{2}}
{(1-4\beta)^{2}}}\
\Big],
\end{equation}
where
\begin{equation}
\alpha(\beta) = {\rm cn}^{-1}\Big(-\frac{1+4\beta}{1-4\beta}\,|\,k^{2}\Big)\ , \ \ \ \ \ \ \ \
\   \beta=\frac{n^{2}}{m^{2}}  \ .
\end{equation}
Since $\beta$ is independent of $\kappa$,
 we can immediately expand at small $k$. Fixing the sign, the result is
\ba
&& \nu_{_F}  =
\frac{2 \pi  n}{m}+\frac{\pi  n (3 m^2+4 n^2)}{2 m (2 n-m) (m+2 n)} k^2
 -\frac{ \pi  n (15 m^6-276 m^4 n^2-304
   m^2 n^4+576 n^6)}{32 m (m-2 n)^3 (m+2 n)^3} k^4 \no  \\
   && -\ \frac{ \pi  n (35 m^{10}-780 m^8 n^2+9696 m^6
   n^4+9856 m^4 n^6-28928 m^2 n^8+25600 n^{10})}{128 m (m-2 n)^5 (m+2 n)^5}k^6+
   ...\label{eq:fermi}
   \ea
It is interesting to understand the singularity of the expansion at $n=\frac{m}{2}$
or
$\beta = \frac{1}{4}$. If we plot the fermionic discriminant as $k$ decreases we can see that there is an antiperiodic solution appearing at $\beta=\frac{1}{4}$
when $k=0$ and being absent for $k>0$. This curious phenomenon is the reason
for  the singularity of the
small $k$ expansion. One can just regulate this ``resonance'' by taking  $m$ to be odd.

\subsubsection{Sum of stability angles and short string expansion of the energy
 }

Let us now combine the above  fluctuation frequencies expanded in powers of
 $\k= { k  m}$
\rf{ioi},\rf{eq:bose} and \rf{eq:fermi}  with proper multiplicities and
signs as they should
appear in the 1-loop correction to the energy in \rf{pou}
\ba
&&  {\nu_{n} = 4\times (\nu_{_{AdS_{5}}}+\nu_{_{S^{5}}})-8\times\nu_{_F} } \no \\
&&
\ \ \ =\frac{4 \pi \k^2 m }{n\,(m^2 -4 n^2)}-\frac{\pi  \k^4
 (2 m^8-28 m^6 n^2+133 m^4 n^4-128 m^2 n^6+48 n^8)}{2m
   n^3 (m^2-4 n^2)^3 (m^2-n^2)} \nonumber \\
   &&\ \ \  +\ \frac{\pi  \k^6  }{16 m^3(m^2-n^2)^2 (m^2
   n-4 n^3)^5}\Big(8 m^{16}-180 m^{14} n^2+1705 m^{12} n^4-8772
   m^{10} n^6+25883 m^8 n^8\nonumber\\
   && \ \ \ \ \ \ \ \ \ \ \ \ \-\ 35456 m^6 n^{10} +25824 m^4 n^{12}-13824 m^2 n^{14}+3840 n^{16}\Big)
   +...\ . \la{summa}
\ea
As a check,
 we observe that the sum over  $n$  of this combination is
  convergent at large $n$.

In the rest of this section we shall
  focus on the case of  $m=1$.
  Dividing by the period $4\K(\kappa^{2})$
we get
\begin{eqnarray}
\frac{\nu_{n}}{\T} &=& \frac{2 }{n-4 n^3}\kappa ^2-\frac{16 n^8-80 n^6+115 n^4-26 n^2+2 }{4 n^3 (n^2-1) (4
   n^2-1)^3}\kappa ^4  \\
   &-& \frac{256 n^{16}-896 n^{14}+2560 n^{12}-5864 n^{10}+5295 n^8-1954 n^6+402 n^4-44 n^2+2 }{8 n^5
   (n^2-1)^2 (4 n^2-1)^5}\kappa ^6+O\left(\kappa ^7\right)\nonumber .
\end{eqnarray}
The sum over modes with $n>1$ then gives
\ba
\frac{ 1}{\T} \sum_{n=2}^{\infty}\ {\nu_{n}} =
\big(\frac{8}{3}-4 \log 2\big)\kappa ^2 &+& \big(\frac{3 \zeta_3}{8}-\frac{347}{432}
+\frac{\log 2}{2}\big)\kappa ^4 \nonumber\\
&+&
   \big(-\frac{63 \zeta_3}{64}-\frac{15 \zeta_5}{64}+\frac{38759}{31104}+
   \frac{\log 2}{4}\big)\kappa ^6 +O\big(\kappa ^8\big) \ .
\la{lpl}
\ea
where we have used the shorthand notation for the Riemann zeta function: $\zeta_k=\zeta(k)$.
 The  $n=0$ contribution  comes only  from $AdS_5$ part and we get
\be
\nu_{0} = 16\,\K(\kappa^{2})\,\kappa \ , \ \ \ \ \  \ \ \ \ \ \ \ \ \ \
\frac{\nu_{0}}{\T} = 4\kappa\ .
\ee
The $n=1$ contribution  comes only from the $AdS_5$  part and the fermions
\be
\frac{\nu_{1}}{\T} = -\frac{5 \kappa ^2}{3}+\frac{401 \kappa ^4}{432}-\frac{18529
 \kappa ^6}{15552}+O\left(\kappa ^8\right) \ .
\ee
Summing up all the contributions   we get
from \rf{pou}    the following expression for the
1-loop correction to the string energy  (taking into account the relation  \rf{repa} valid
in the static gauge $t=\k \tau$)
\ba
\mathcal E_1=
\frac{1}{2 \T \k } \sum_{n=-\infty}^{\infty} {\nu_{n}} &=& 2 +\kappa  (1-4 \log 2)+ \frac{1}{8}
\kappa ^3
\Big({3 \zeta_3}+ {1}+4 \log 2 \Big) \nonumber \\
&&+ \frac{ 1}{4} \kappa ^5 \Big(-\frac{63 \zeta_3
   }{16}-\frac{15 \zeta_5}{16}+\frac{7}{32}+ {\log 2}\Big)+O(\kappa
    ^7)   \ . \la{ghjk}
\ea

In general, we can organize the short string expansion of the energy as
\be
&&E = E\Big(\frac{N}{\sqrt\lambda}, \sqrt\lambda\Big) = \sqrt\lambda\,\mathcal E_{0}(\mathcal N)+
\mathcal E_{1}(\mathcal N)+\frac{1}{\sqrt\lambda}\,\mathcal E_{2}(\mathcal N)+... \  ,
\la{kouu}\\
&&
\mathcal E_{k}= \sqrt{2\mathcal N}\ \Big(a_{0k}+a_{1k}\,\mathcal N+a_{2k}\,\mathcal
N^{2}+... \Big)
 + c_{0k}+c_{1k}\,\mathcal N+.... \ . \la{jrkp}
\ea
where $c_{nk}$ are coefficients of  ``non-analytic''
 terms \cite{rt}.
Using \rf{nen},\rf{ses}  and \rf{ghjk} we thus find that
for the pulsating string in  $\mathbb{R}\times S^{2}$
\ba
\mathcal E_{0} &=& \sqrt{2\mathcal N}\Big(1-\frac{1}{8}\mathcal N-\frac{5}{128}\mathcal N^{2}
+... \Big), \la{kla}\\
E_1\equiv \mathcal E_{1} &=& 2+\sqrt{2\mathcal N}\Big[
 1-4 \log 2+\Big(\frac{3}{2}\log 2+\frac{3}{4} \zeta_{3}+\frac{1}{8}\Big)\mathcal N
  \no \\
 &&  + \  \Big(
\frac{25}{32}\log 2-\frac{135}{32}\zeta_{3}-\frac{15}{16}\zeta_{5}+\frac{11}{128}
\Big)\mathcal N^{2}+...\Big]. \la{kqq}
\ea
The energy can be re-written
in terms of $N$ and the string tension as follows
\ba
&&E = \sqrt{2N\sqrt\lambda}\ \Big(a_{00}+\frac{a_{10}N+a_{01}}{\sqrt\lambda}+
...\Big)  + c_{01}+ ... \ , \la{ert} \\
&&a_{00}=1, \qquad a_{10}=-\frac{1}{8}, \qquad a_{01} = 1-4\log 2, \qquad c_{01}=2\ ,\ \ \  ...
\la{yyyi}
\ea

\subsection{Pulsating string in $AdS_3$}

The aim of this  subsection is to  use the results of section 3 to
compute,  in a similar way as above,
 the one-loop correction to the energy in the
short string limit of small oscillation  parameter
$\mathcal N\to 0$ or small classical energy $\E_0\to 0$.


In contrast to the pulsating string in $\mathbb{R}\times S^{2}$  where we can  use the
 static gauge  $t= \k \tau$
  in which the relation between the 2d and space-time energy is simple,
 here  this is no longer the case
 as in the conformal gauge
 the classical solution for the
  $AdS_5$   time $t$  depends on the  world-sheet time
  $\tau$ in a  non-linear way.
Here we may  fix the static gauge on the fluctuation of $t$
(i.e. set it to zero)
while using the  classical conformal-gauge  relation between $t$ and $\tau$ in \rf{glo},
i.e.\foot{For comparison, the analog of the classical
 energy parameter ${\mathcal E}$
in the case of puslating string on $S^2$  in the
previous subsection where we had $dt = \k d \tau$
was $\kappa$.}
\be
\label{eq:t-tau}
dt = \dot t\ d\tau = \frac{\mathcal E_0}{\cosh^{2}\rho(\tau)}\,d\tau\ .
\ee
Since the relation between $t$ and $\tau$ is  a change of variable,
it does not matter for  the equations of motions
and the fluctuation operator which can be solved  in terms of the $\tau$ variable.
What is affected  is the expression for  the period,
which for the  $t$-motion is then
\be
\T ={\mathcal E_0} \int_{0}^{\T_{\tau}}\frac{d \tau }{\cosh^{2}\rho(\tau)}\ ,\la{ptre}
\ee
where $\T_{\tau}$ is the period in the $\tau$ variable.
Having found the stability angles we should then
use again the expression \rf{pou} where to get the space-time
 energy we will need to divide
by the period $\T$ in \rf{ptre} corresponding to the variable $t$.


\subsubsection{Stability angles}
As follows from section 3 the  bosonic  fluctuations of type I
obey the equation
\be
\label{PulsAdSFluctI}\mathcal O_{I}\,\zeta_{n} \equiv \left[
-\partial_{\tau}^{2}-n^{2}-z\sinh^{2}\rho
\right]\,\zeta_{n} = 0,
\ee
where $z=2$ for the non-trivial boson, and $z=0$ for the free modes.
The stability angle for $\mathcal{O}_{I}$ is
\be
\nu_{I} = \pm 4\ \K \left(i\,\Z(\alpha\,|\,k^{2})+\frac{\pi}{2 \K}\right) \equiv \pm
4\,\K \,i\,\Z(\alpha\,|\,k^{2}),
\ee
where ($w$ is defined in \rf{xerx})
\be
{\rm sn}(\alpha\,|\,k^{2})  = \frac{1}{k}\sqrt{1-k^2-\frac{n^{2}}{w^{2}}}.
\ee
Expanding in the limit $\mathcal{E}_0\to 0$  (cf. \rf{e4})
\be \la{kiy}
\nu = \nu^{(1)} {\mathcal E}_0^{2} + \nu^{(2)} {\mathcal E}_0^{4} +\nu^{(3)} {\mathcal E}_0^{6} +... \  , \ee
 we find  for $|n|\neq 1$ (see Appendix~D)
\ba
\label{PulsAdSStabI}
\nu_{I,n}^{(1)} &=& \pi\frac{3n^{2}-z}{2n}, \nonumber \\
\nu_{I,n}^{(2)} &=& -\pi\,\frac{105 n^6-15 n^4 (2 z+7)-3 n^2 (z-10) z+2
z^2}{32 n^3 \left(n^2-1\right)}, \la{rgo}
\\
\nu_{I,n}^{(3)} &=& \frac{\pi  \left(1155 \left(n^2-1\right)^2 n^6-
\left(5 n^4-5 n^2+2\right) z^3- \left(35 n^4-65 n^2+24\right) n^2 z^2-315
   \left(n^2-1\right)^2 n^4 z\right)}{128 n^5 \left(n^2-1\right)^2}.\nonumber
\ea
The singularity at $n=\pm 1$ is  absent for $z=0$. In the $z=2$ case, a more detailed analysis
shows that at the considered orders we have  $\nu_{I,\pm 1}^{(2)}=0$.

In the $z=2$ case, there is a singularity at $n=0$. At leading order the problem is related to the fact that the
stability angle $\nu_{0}$ of the equation
\be
(-\partial_{y}^{2}-2\mathcal E_0^{2}\sin^{2}y)\,\zeta(y)=0,\qquad \zeta(a+2\pi) = e^{i\nu_{0}}\zeta(a),
\ee
goes like $\mathcal E_0$ for $\mathcal E_0\to 0$. This is valid in general and can be checked numerically in the present
case. The precise result in our case is
\be
\nu_{I,0} = 2\pi\mathcal E_0 + \dots.
\ee

In the case of type II  fluctuation equation we have  for $m=1$
\be
\mathcal O_{II}\,\zeta_{n} \equiv \Big[
-\partial_{\tau}^{2}-n^{2}+\frac{2\mathcal E_0^{2}}{\sinh^{2}\rho}-2\sinh^{2}\rho
\Big]\,\zeta_{n} = 0,
\ee
where
\be
\sinh\rho(\tau) = \sqrt{\frac{-R_{+}R_{-}}{R_{+}-R_{-}}}\,{\rm sd}\Big(\sqrt{R_{+}-R_{-}}\,\tau
\,|\, \frac{R_{+}}{R_{+}-R_{-}}\Big).
\ee
As we have argued in section 3.3  this operator is of the Lam\'e type.
The stability angle is then
\be
\nu_{II} = \pm 4\ \K(k^{2}) \Big(i\,\Z(\alpha\,|\,p^{2})+\frac{\pi}{2 \K(p^2)}\Big),
\ee
where ($B$ and $p$ were  defined in \rf{prt},\rf{ltl})
\begin{equation}
{\rm sn}(\alpha\,|\,p^{2})  = \sqrt{\frac{1+p^2-B}{p^2}}.
\end{equation}
The coefficients in the small ${\mathcal E}$
 expansion of the stability angle are explicitly
\ba
\label{PulsAdSStabII}
\nu_{II, n}^{(1)} &=& \pi\frac{n \big(3 n^2-7\big)}{2 (n^{2}-1)}, \nonumber \\
\nu_{II, n}^{(2)} &=& -\pi\,\frac{n \big(105 n^6-435 n^4+603 n^2-337\big)}{32
\big(n^2-1\big)^3}, \la{ergo} \\
\nu_{II, n}^{(3)} &=& -\frac{\pi  n \big(1155 n^{10}-7035 n^8+17150 n^6-21430
 n^4+14159 n^2-4511\big)}{128 \big(n^2-1\big)^5}.\nonumber
\ea
The same result can
 be obtained without knowing the analytical expression for  $\nu_{II}$
and using  perturbation theory alone.
 The detailed calculation is reported for completeness in Appendix ~D.


The type III (fermionic)  fluctuation equation  has the form
\be
\mathcal O_{III}\,\zeta_{n} \equiv \Big[
-\partial_{\tau}^{2}-n^{2}-\sinh^{2}\rho\pm i\,\frac{d}{d\tau}\sinh\rho
\Big]\,\zeta_{n} = 0.
\ee
The stability angle for the fermionic operator $\mathcal{O}_{III}$ is
\ba
&&\nu_{III} = \pm 4\ i\ \mathbb{K}\,\Big[
\frac{1}{2}\,Z(\alpha({\beta})\,|\,k^{2})+
i\,\sqrt{\beta}\,\sqrt{1+\frac{16\,\beta\,k^{2}}
{(1-4\beta)^{2}}}
\Big],\\
&&
\alpha(\beta) = {\rm cn}^{-1}\Big(-\frac{1+4\beta}{1-4\beta}\,|\,k^{2}\Big),\ \ \ \ \ \ \
\
\beta=\frac{n^{2}}{w^{2}} \ . \ea
Expanding in small $\mathcal{E}$ we obtain
the stability angle \rf{kiy} with coefficients
\ba
\nu_{n}^{(1)} &=& \pi\,\frac{n\,(12n^{2}-7)}{2(4n^{2}-1)}, \nonumber \\
\nu_{n}^{(2)} &=& -\pi\frac{n (6720 n^6-6960 n^4+2412 n^2-337
)}{32 (4 n^2-1)^3}, \la{ferg} \\
\nu_{n}^{(3)} &=&
 \frac{\pi  n \big(1182720 n^{10}-
 1800960 n^8+1097600 n^6-342880 n^4+56636 n^2-4511\big)}{128 (4 n^2-1)^5}
 \nonumber.
\ea

\subsubsection{Sum of stability angles and short string expansion of the energy
 }

Adding together the contributions of the
 $AdS_{5}$ and $S^{5}$ bosonic modes (including the 5 massless modes with $z=0$)
 and the fermions in \rf{rgo},\rf{ergo},\rf{ferg}   we obtain for the  small $\mathcal E_0$
 expansion of the sum of individual    stability  angles
(for $n\ge 2$)
\ba
&&\frac{1}{\pi}\nu_{n}
 = 5\times \Big(\frac{3n^{2}-0}{2n}\mathcal E_0^{2}+ \dots\Big)+
1\times \Big(\frac{n (3 n^2-7)}{2 (n-1) (n+1)}\mathcal E_0^{2}+\dots\Big)+
  2\times \Big(\frac{3n^{2}-2}{2n} \mathcal E_0^{2}+\dots\Big)
\nonumber \\
&&\qquad \qquad
-\ 8\times\Big( \frac{n(12n^{2}-7)}{2n(4n^{2}-1)}\mathcal E_0^{2}+\dots\Big) \la{potr} \\
&&
=-\ \frac{2\pi \, (2 n^2+1)}{n(n^{2}-1)(4n^{2}-1)}\mathcal E_0^{2}+\frac{\pi(240 n^{12}-560
   n^{10}+713 n^8-361 n^6+83 n^4-8 n^2+1)}{2 n^{3}(n^{2}-1)^3 (4n^{2}-1)^3 }
   \mathcal E_0^{4}+\dots\nonumber~.
   \ea
The $n=\pm 1$ contributions come  from fermions and free bosonic modes
\be
2\nu_{1} = \frac{5 \pi }{3}\mathcal E_0^{2}+\frac{505 \pi  }{432}\mathcal E_0^{4}
-\frac{105515}{31104}\,\mathcal E_0^{6}+\dots\ .
\ee
Then the total sum for $n \not=0$ is
\ba
\sum_{n\neq 0}^{\infty}\nu_{n} &=& \pi \big(-10+16\log 2\big)\mathcal E_0^{2}+\pi\Big(
\frac{199}{8}-30\log 2-\frac{3}{2}\zeta_{3}\Big)\mathcal E_0^{4}b \no \\
&&
+\ \Big(
-\frac{9395\pi}{128}+\frac{315\pi}{4}\log 2+\frac{111\pi}{16}\zeta_{3}+\frac{15}{16}\zeta_{5}
\Big)\,\mathcal E_0^{6}+
\dots\la{fwq}
\ea
The overall (negative)
sign with which this sum enters the expression for the 1-loop energy
can be
 fixed by looking at the contribution of the
 free $S^{5}$ modes. The first correction
to the period in the $\tau$ variable is negative
\be
\T_{\tau} = \frac{4\mathbb K(\frac{R_{+}}{R_{+}-R_{-}})}{\sqrt{R_{+}-R_{-}}} =
2\pi-\frac{3\pi}{2}\mathcal E_0^{2}+\frac{105\pi}{32}\mathcal E_0^{4}+\dots.
\ee
Adding the zero mode $n=0$  contribution of the non-trivial type I fluctuation
\rf{PulsAdSFluctI}
(multiplied by 2 which is the number of   bosons with $z=2$) we find
\be
E_{1} = \frac{1}{2\T}\Big(2\cdot 2\pi\mathcal E_0-\sum_{n\neq
0}^{\infty}\nu_{n} \Big) ,
\ee
where $\T$ is the period of  the $t$ variable in  \rf{ptre}
\be
\T =\mathcal E_0  \int_0^{\T_\tau}  \frac{d \tau }{\cosh^{2}\rho}
 = 2\pi\mathcal E_0-\frac{5}{2}\pi\mathcal E_0^{3}+\frac{189\pi}{32}\mathcal E_0^{5}+\dots.
\ee
Using the fact that the classical energy parameter
is related to the oscillation number as in \rf{eeee}  (here  $m=1$)
\be
\mathcal E_0 = \sqrt{2\mathcal N}\ \Big(1+\frac{5}{8}\mathcal
N-\frac{77}{128}\mathcal N^{2}+\dots\Big)\ ,
\ee
we finally obtain (cf. \rf{kqq})
\begin{eqnarray}
&& E_{1}=  1 + \sqrt{2\mathcal N}\ \Big[\frac{5}{2}-4\log 2+\Big(
-\frac{37}{8}+\frac{5}{2}\log 2+\frac{3}{4}\zeta_{3}
\Big)\,\mathcal N \nonumber\\
&& \qquad \qquad \qquad\ \ \ \ \ \ \ \ + \  \Big(
\frac{3915}{256}-\frac{231}{32}\log 2-\frac{117}{32}\zeta_{3}-\frac{15}{16}\zeta_{5}\Big)\,
\mathcal N^{2}+\dots\Big]\ .\la{trq}
\end{eqnarray}

\renewcommand{\theequation}{6.\arabic{equation}}
 \setcounter{equation}{0}
\setcounter{section}{5}

\section{One-loop correction to energy of folded string
in $\mathbb{R} \times S^2$
}

To get a  better  understanding of the  structure of energy of ``small' semiclassical
strings it is useful to supplement the discussion
of the folded spinning string in $AdS_3$ in \cite{tt,bd}
and the analysis  of the pulsating strings  in $\mathbb{R} \times S^2$
and $AdS_3$   carried out above   with a   similar study
of the  1-loop corrected energy of spinning folded  string in
$\mathbb{R} \times S^2$
part of $\ads$. This  will be the aim of this section.

  We shall start with the case of the folded string in $S^3$
   moving  along  big circle   with orbital momentum
   $J_1=\sql \J_1$   and  spinning around its c.o.m. with momentum $J_2=\sql \J_2$.
  When discussing  one-loop corrections we will eventually specify
  to the case of $\J_1=0$ and  expand in $\J_2 \to 0$.


\subsection{Classical solution}

Let us start with a brief review of the
  folded string with two angular momenta
   moving in $S^3\subset \ads$ \cite{ftp}. The metric of  $\mathbb{R} \times S^3$ is
with metric
\be
ds^2 = -dt^2+d\theta^2+\cos^2\theta\,d\varphi_1^2+\sin^2\theta\,d\varphi_2^2.
\ee
and the ansatz  one assumes is ($i=1,2$)
\be
t = \kappa\,\tau,\qquad \theta=\theta(\sigma), \qquad \varphi_i = w_i\,\tau  \  .
\ee
In conformal gauge the  only non-trivial equation of motion reads (we assume $w_2>w_1$)
\be
\theta'' + \frac{1}{2}\,w_{21}^2\,\sin(2\theta) = 0,\qquad w_{21}^2 = w_2^2-w_1^2 \ ,
\ee
which has   the Virasoro condition as its first integral
\be
\theta'^2 + w_1^2\,\cos^2\theta + w_2^2\,\sin^2\theta = \kappa^2 \ .\la{klop}
\ee
The periodic solution with $\theta(0) = 0$ is  (see also  \cite{bfst,mtt06})
\be
\sin\theta = \sqrt{q}\,{\rm sn}(w_{21}\,\sigma | q),\qquad\cos\theta = {\rm dn}(w_{21}\,\sigma | q),
\ee
where
\ba
q = \sin^2\theta_0 = \frac{\kappa^2-w_1^2}{w_2^2-w_1^2}, \qquad
w_{21} = \sqrt{w_2^2-w_1^2} = \frac{2}{\pi}\mathbb{K}(q).
\ea
Let us note that the equation \rf{klop} can
 be written in a form which depends only on $q$
\be
\theta'^2 = w_{21}^2\,(\sin^2\theta_0-\sin^2\theta) =
\Big[\frac{2}{\pi}\mathbb{K}(q)\Big]^2\,(q-\sin^2\theta) \ ,\qquad \theta(0) = 0.
\ee
The expressions for  the energy and the two angular momenta
 can be given, e.g.,  in terms of the
hypergeometric functions
\ba
\mathcal E_0   = \kappa, \qquad
\J_1 =\frac{w_1}{w_{21}}\ {}_2F_1\Big(-\frac{1}{2}, \frac{1}{2}, 1, q\Big), \qquad
\J_2 = \frac{w_2}{w_{21}}\,\frac{q}{2}\  {}_2F_1\Big(\frac{1}{2}, \frac{3}{2}, 2, q\Big).
\ea
$\J_i$  satisfy the relationship
\be
\frac{\J_1}{w_1} + \frac{\J_2}{w_2} = 1.
\ee
Useful  relations which allows to eliminate $q$
and find $\E_0=\E_0(\J_1,\J_2)$
is \cite{bfst}
\ba
\Big(\frac{\mathcal E_0}{\mathbb{K}(q)}\Big)^2 -\Big(\frac{\J_1}{\mathbb{E}(q)}\Big)^2 = \frac{4}{\pi^2}\,q, \qquad
 \Big(\frac{\J_2}{\mathbb{K}(q)-\mathbb{E}(q)}\Big)^2 -\Big(\frac{\J_1}{\mathbb{E}(q)}\Big)^2 = \frac{4}{\pi^2}.
 \label{j2}
\ea
In the short string limit, i.e. the small $q$ limit,
 the solution for $\theta$  can be expanded as
\begin{equation}\la{quki}
\theta(\s) = \sqrt{q}\sin \sigma +\Big(\frac{3 \sin \sigma}{16}+\frac{\sin (3 \sigma )}{48} \Big) q^{3/2}
 + \Big(\frac{23 \sin \sigma}{256}+\frac{\sin (3
   \sigma )}{64} +\frac{\sin (5 \sigma )}{1280}\Big) q^{5/2}+O\Big(q^{7/2}\Big).
\end{equation}
To expand the energy in  small   spins  (or $q\to 0$)  one may consider two special
scaling limits.
The first is  when
\be
\J_2 \to 0 \ , \ \ \ \ \ \ \ \   r= \frac{ \J_1}{ \J_2}={\rm fixed} \ . \la{raf}
\ee
Then
\ba
\J_2 &=& \frac{q}{2}+\frac{3 q^2}{16}+\frac{8 r^2+15}{128} q^3+\frac{208 r^2+175 }{2048}q^4+O(q^5), \nonumber\\
w_1 &=& \frac{r}{2}q+\frac{7 r}{16}q^2+\frac{r(8 r^2+47)}{128}
 q^3+\frac{r(272 r^2+639) }{2048}q^4+O(q^5),\no  \\
w_2 &=& 1+\frac{q}{4}+\frac{8 r^2+9}{64} q^2+\frac{48 r^2 +25}{256}
 q^3+\frac{64 (6 r^2+55) r^2+1225
   }{16384}q^4+O(q^5), \nonumber\\
\kappa &=& \sqrt{q}+\frac{r^2+2}{8} q^{3/2}-\frac{r^4-24 r^2 -18 }{128} q^{5/2}+\frac{
r^6+10 r^4+220 r^2+100
   }{1024}q^{7/2}+O(q^{9/2}) \la{popi}
\ea
Expressing the  classical energy in terms of
 $\J_2$  we get
\ba
\E_0 = \kappa = \sqrt{2\,\J_2}\ \Big(1+\frac{2 r^2+1}{8} \J_2-\frac{4 r^4-28 r^2-3}{128}  \J_2^2+
  \frac{8 r^6-52 r^4+94 r^2+1
   }{1024}\J_2^3+ ...  \Big)\ .\la{iopi}
\ea
Another option is
\be
\J_2 \to 0 \ , \ \ \ \ \ \ \ \   s = \frac{ \J_1^2  }{ \J_2}={\rm fixed} \ . \la{saf}
\ee
In this case, expanding in small $q$, we find
\ba
\J_2 &=& \frac{q}{2}+\frac{2 s+3}{16}  q^2+\frac{2 s(s+10)+15}{128}  q^3
+\frac{4 s (17 s+81)+175 }{2048}q^4+O(q^5), \nonumber \\
w_1 &=& \frac{\sqrt{r} }{\sqrt{2}}\sqrt{q}+\frac{\sqrt{s} (2 s+11) }{16
 \sqrt{2}}q^{3/2}+\frac{\sqrt{s} (4 s (s+25)+259) }{512 \sqrt{2}}q^{5/2}+O(q^{7/2}),
 \nonumber \\
w_2 &=& 1+\frac{s+1}{4} q+\frac{2 s(s+9)+9}{64}  q^2+\frac{4 s(4s+17)+25}{256}  q^3\no \\
&& \ \ \ \ \ \ + \frac{4 s (-2 s^3+352 s+985)+1225
   }{16384}q^4+O(q^5) \ ,  \la{koq}
\\
\kappa &=& \sqrt{\frac{s}{2}+1} \ q^{1/2} +\frac{s (2 s+11)+8 }{16 \sqrt{2} \sqrt{s+2}}q^{3/2}+\frac{ 4s^4  29s^3+515s^2 +760s +288 }{512 \sqrt{2}
   (s+2)^{3/2}}q^{5/2}+O(q^{7/2}).\no
\ea
Then the  classical energy is (cf. \rf{iopi})
\ba
&& \E_0 =\kappa =\sqrt{ (2 +s) \J_2}\ \Big(
1+\frac{2 s+1 }{4 s+8}\J_2-\frac{4 s^3+4 s^2-14 s-3 }{32 (s+2)^2}\J_2^2\no \\
&& \qquad \qquad\qquad \qquad \ \ \ \ \ \ +\
 \frac{8 s^5+8 s^4-80 s^3-135 s^2+1 }{128
   (s+2)^3}\J_2^3+...  \Big)\  . \la{tytr}
\ea

\subsection{Quadratic fluctuation operators}

The bosonic fluctuation Lagrangian near this
solution  was found
in conformal gauge in \cite{mtt06}.
In $AdS_5$ we have one massless mode and four modes
 with $M^2 = \kappa^2$  while for the $S^5$  fluctuations we get
\ba
 \tilde{L}_{_{S^5}}
  &=& |\dot X|^2-|X'|^2-M_X^2\,|X|^2 + \frac{1}{2}(\dot \eta^2-\eta'^2-M_\eta^2\,\eta^2) +
  (Q_1\,f_1 + Q_2\,f_2)\,\dot \eta\no  \\
&& \ \ \  +\  \frac{1}{2}(\dot f_1^2-f_1'^2-M_1^2\,f_1^2) +
\frac{1}{2}(\dot f_2^2-f_2'^2-M_2^2\,f_1^2) \ ,
 \la{flii}\\
\label{eq:mx2}
M_X^2    &=& 2\,(\kappa^2-w_1^2)\,\frac{\sin^2\theta}{\sin^2\theta_0}+2\,w_1^2-\kappa^2, \quad
M_\eta^2 =-(\kappa^2-w_1^2)\,\frac{\cos 2\theta}{\sin^2\theta_0}, \\
M_1^2    &=& -(\kappa^2-w_1^2)\,\Big(1-2\,\frac{\sin^2\theta}{\sin^2\theta_0}\Big), \quad
M_2^2    =-(\kappa^2-w_1^2)\,\Big(1+\frac{\cos 2\theta}{\sin^2\theta_0}\Big), \\
Q_1 &=& \phantom{-}2\,w_1\,\sin\theta, \quad \quad
\label{eq:omega2}
Q_2 = -2\,w_2\,\cos\theta.
\ea
We observe that when  both  spins are non-trivial
there are  three coupled bosonic fluctuations and this   makes the
exact
 computation of the fluctuation determinant a  non-trivial task.

  Below we shall  consider
a particular case
with only  one non-zero spin $\J_2$ (the one corresponding to  rotation  around c.o.m.).
In this case the bosonic fluctuations can be decoupled in the static
gauge.
We  shall  thus set\footnote{Let us note that the condition $\kappa < w $ implies an upper bound for $\mathcal{J}_2$.}
\begin{equation}
w_1=0, \quad \quad w \equiv w_2, \quad \quad \mathcal{J}_1=0 \ .
\end{equation}
In this case the expansions of the classical energy in \rf{iopi} and \rf{tytr} become the same ($r=s=0$)
\ba
\E_0 = \kappa = \sqrt{2\,\J_2}\ \Big(1+\frac{1}{8} \J_2  +  \frac{3}{128}  \J_2^2+
  \frac{1
   }{1024}\J_2^3+ ...  \Big)\ .\la{ioupi}
\ea
The quadratic bosonic fluctuation action  in the static gauge  is
found to be (see Appendix C; here $k=1,2,3,4; \ i=1,2,3$)
\begin{eqnarray}
\tilde{L}&=&\frac{1}{2}  \Big[(\partial_{\sigma}
 \tilde{\eta}_k)^2-(\partial_{\tau} \tilde{\eta}_k)^2+ \kappa^2 \tilde{\eta}_k^2+
 (\partial_{\sigma} \tilde{\psi}_i)^2-(\partial_{\tau} \tilde{\psi}_i)^2+ (2 w^2 \sin^2 \theta
 -\kappa^2) \tilde{\psi}_i^2 \nonumber\\
&&\ \ \   +  (\partial_{\sigma} f)^2-(\partial_{\tau} f)^2  + f^2 \kappa^2 (1- \frac{2 (\kappa^2
-w^2)}{\theta'^2})\Big]. \la{stata}
\end{eqnarray}
The  fermionic fluctuation operator  is  (here $s_1={\rm sign} \ \theta'$, \ see  Appendix C)
\be
D_F = s_1\,\Gamma_0\partial_{\tau}-\Gamma_7\partial_{\sigma}+u\,\Gamma_{078}\Gamma_{1234} \ ,
\ee
with the corresponding squared operator whose determinant gives fermionic contribution to
1-loop energy
being
\be
D_{F\pm}^2=\partial_{\tau}^2-\partial_{\sigma}^2+u^2 \pm \,u' \ , \ \ \ \ \ \ \ \ \ \ \
u\equiv w \sin \theta \ . \la{fesa}
\ee
The UV finiteness in the static gauge is  checked as follows (cf. \rf{rwrr}):
the sum of (mass)$^2$ terms
\begin{eqnarray}
AdS &:& 4\times \kappa^{2}, \nonumber\\
S^{5} &:& 3\times (2 w^2 \sin^2 \theta - \kappa^2), \nonumber\\
&& 1\times \kappa^{2}\,(1-\frac{2 (\kappa^2-w^2)}{\theta'^2}), \nonumber\\
F &:& -8\times w^2 \sin^2 \theta,
\end{eqnarray}
gives
\begin{equation}
\frac{2}{\theta'^2}[\theta'^4 - \kappa^2 (\kappa^2-w^2)] = \sqrt{-g}\,R^{(2)},
\end{equation}
which is the  expected value for  UV finiteness in the static gauge  (cf. \rf{rwwr}).

Since there is no nontrivial dependence of the potentials on $\tau$, we may
 switch to Euclidian time $\tau  \rightarrow i \tau$ and
 replace $\partial_{\tau} \rightarrow i \omega$ . Then the  relevant  1-d
operators will have the form
\begin{equation}
\mathcal{O}= - \partial_{\sigma}^2 + M^2(\s) + \omega^2 \ ,
\end{equation}
and can be put as in \ci{bd}  in the Lam\'e form allowing us to compute the determinants in a closed form.
For the bosonic fluctuation in \rf{stata} with
 mass $M^2= 2 w^2 \sin^2 \theta -\kappa^2$ we obtain the operator
\begin{equation}
\mathcal{O}_{I} = w^2\,\Big[-\partial_x^2 + 2 k^2 {\rm sn}^2(x| k^2) - \frac{\kappa^2 - \omega^2}{w^2}\Big]
\ ,\ \ \ \ \ \ \ \ \
k^2 \equiv \frac{\kappa^2}{w^2}, \quad \quad x= w \sigma. \la{opac}
\end{equation}
For the fluctuation with  mass $M^2= \kappa^2 (1- 2 \frac{\kappa^2-w^2}{\theta'^2})$ we get  a similar result
\begin{equation}
\mathcal{O}_{II} = w^2\,\Big[-\partial_{\bar{x}}^2 + 2 k^2 {\rm sn}^2(\bar{x}| k^2) - \frac{\kappa^2 - \omega^2}{w^2}\Big]
\ , \ \ \ \ \ \ \ \ \ \    \bar{x}\equiv  x + i \K'+\K      \ ,  \end{equation}
where  $\K'=\K(q'), \  \ q'= \sqrt{1-q}$.
The fermionic operator in \rf{fesa}  can be written as
\begin{equation}
\mathcal{O}_{III}=-\partial_{\sigma}^2+\omega^2 + q w^2 {\rm sn}^2(w \sigma|q)
 \pm w^2 \sqrt{q}\,{\rm cn}(w \sigma|q) \,{\rm  dn}(w \sigma| q) ,\label{fs2}
\end{equation}
which  can also  be put in Lam\'e form as in  \cite{bd} for both signs in the potential
(here we ignore  an irrelevant  overall constant factor)
\ba
\mathcal{O}_{III}=-\partial_x^2 + 2 \tilde{q} \ {\rm sn}^2(x| \tilde{q})+ \bar{\omega}^2,
\ea
where
\ba
x=\left\{
    \begin{array}{ll}
      \frac{\tilde{\mathbb{K}}}{\pi}\sigma, & \mbox{for} \ + \mbox{sign}\\
      \frac{\tilde{\mathbb{K}}}{\pi}\sigma+\tilde{\mathbb{K}}, & \mbox{for} \ - \mbox{sign}
    \end{array}
  \right. , \qquad \tilde{q}=\frac{4 \sqrt{q}}{(1+\sqrt{q})^2}, \qquad \bar{\omega}^2= (\frac{\pi \omega}{\tilde{\mathbb{K}}})^2+\tilde{q} \ , \ \ \ \ \tilde{\mathbb{K}}\equiv \mathbb{K}(\tilde{q}) \ .
\ea

\subsection{Short string expansion of the one-loop energy}

As in the case of the folded spinning string in $AdS_3$ the one loop correction to the energy is given simply
by \be \la{energ}
E_1 = \frac{1 }{ \k}  E_{2d}
 \ , \ \ \ \ \ \ \ \ \ \ \ \ \   E_{2d}= \frac{\Gamma_1}{{\cal T}}
 \ , \la{tyr}\ee
where $\Gamma_1$ is the one-loop  Euclidean effective action for all fluctuations in the static gauge
and $\T$ is an arbitrary infinite time interval.
 Below we will consider the expansion of $E_1$  in  the short string
(small spin) limit, i.e. expand the determinants in $\Gamma_1$ in the limit  $q\to 0$ (cf. \rf{quki}).
We shall not use exact expressions for the determinants as in \ci{bd}
but rather apply direct perturbation theory in small $q$.
 For small $q$, the 1d fluctuation operators are ($\kappa = \sqrt q + \dots$)
\ba
\mathcal O_{AdS_{5}} &=& -\partial_{\sigma}^{2}+\omega^{2}+q+\dots,\nonumber \\
\mathcal O_{S^5} &=&\mathcal O_{I,II}  = -\partial_{\sigma}^{2}+\omega^{2}+q\,(2\sin^{2}\sigma-1)+\dots, \\
\mathcal O_{F} &=&  \mathcal O_{III} = -\partial_{\sigma}^{2}+\omega^{2}+q\,\sin^{2}\sigma\pm\sqrt{q}\,\cos\sigma+\dots
\quad .\nonumber
\ea
Using the standard  ``quantum mechanical''  perturbation theory  in
 the basis $\langle\sigma |n\rangle = \frac{1}{\sqrt{2\pi}}e^{in\sigma}$  we  find the leading terms in the spectra\foot{Here the $S^{5}$  mode  at $n=1$ requires diagonalization of a $2\times 2$ matrix, or the change to the basis
$\sin\sigma, \cos\sigma$. We formally  associate the two corresponding  eigenvalues to $n=\pm 1$.}
\ba
&&\Omega^2= \omega^2 +  \omega_n^2 \ , \ \ \ \ \ \ \ \  \ \ \ \ \
\omega_{n, AdS_{5}}^{2} = n^{2} + q + \dots, \nonumber\\
&& \omega_{n, S^{5}}^{2} =n^{2} + 0 + \dots\qquad |n|\neq 1,  \ \ \ \ \ \ \ \ \ \ \
\omega_{\pm 1, S^{5}}^{2} = 1  \pm \frac{q}{2} + \dots, \la{swqwe} \\
&&\omega_{n, F}^{2} = n^{2} + q\Big(\frac{1}{2}+ \frac{1}{2(4n^{2}-1)}\Big) + \dots
\quad .
\nonumber
\ea
Here the  fermionic  contribution is found by  combining  the  first-order perturbation term
 $\langle\sin^{2}\sigma\rangle = \frac{1}{2}$  with the  second-order one
order piece
\be
\sum_{m\neq n}\frac{|\langle n | \sqrt q \cos\sigma | m\rangle|^{2}}{n^{2}-m^{2}} = \frac{q}{2}\frac{1}{4n^{2}-1} \ .
\ee
The resulting expression for the 1-loop effective action or 2d energy  is then
\ba
E_{2d}  &=& \frac{1}{4\pi}\int_{-\infty}^{\infty}d\omega\sum_{n=-\infty}^{\infty}\log\frac{(n^{2}+\omega^{2}+    \omega_{n, S^{5}}^{2}        )^{4}  \  (n^{2}+\omega^2 +
	\omega_{n, AdS_{5}}^{2})^{4}}{(n^{2}
+\omega^{2}
+ \omega_{n, F}^{2}   )^{8}}  \no   \\
  &=&
 \frac{1}{2}\sum_{n=-\infty}^{\infty}\Big(
	4\omega_{n, AdS_{5}}+4\omega_{n, S^{5}}-8\omega_{n, F} \Big) \ . \la{qqrr}
\ea
The exact  0-mode ($n=0$) contribution here is $2 \kappa$.
Summing up the $n \not=0$ contributions gives
\be
	E_{2d} = 2\sqrt q + q (2-4\log 2)+ O(q^2)  \  . \la{ytyr}
\ee
This  computation can be extended to higher orders in $q$. We find for the
expansion of the  operators
\ba
\mathcal O_{AdS_{5}} &=& -\partial_{\sigma}^{2}+\omega^{2}+q+\frac{q^{2}}{2}+\frac{11}{32}q^{3}+\dots,
\nonumber\\
\mathcal O_{S^5} &=& -\partial_{\sigma}^{2}+\omega^{2}-q\,\cos2\sigma+\frac{q^{2}}{2}[-1+
(3+\cos 2\sigma)\sin^{2}\sigma]  \nonumber\\
&& + \frac{q^{3}}{256}(32-85\cos 2\sigma-32\cos 4\sigma-3\cos 6\sigma) +\dots,  \\
\mathcal O_{F} &=& -\partial_{\sigma}^{2}+\omega^{2}\pm  {q}^{1/2} \,\cos\sigma+q\,\sin^{2}\sigma
\pm \frac{q^{3/2}}{16}(5\cos\sigma+3\cos 3\sigma) \no  \\
&& + \frac{q^{2}}{4}(3+\cos 2\sigma)\sin^{2}\sigma+\frac{q^{5/2}}{256}
(47\cos\sigma+36\cos 3\sigma+5\cos 5\sigma)\nonumber \\
&& +\frac{q^3}{128}(79+38\cos 2\sigma+3\cos 4\sigma)\sin^{2}\sigma +\dots\quad ,    \nonumber
\ea
and the  perturbation theory  then gives the expansion of the  spectra
\ba
\label{eq:foldedS-frequencies}
\omega_{n, AdS_{5}}^{2} &=& n^{2}+ q + \frac{q^{2}}{2}+\frac{11}{32}q^{3}+\dots, \nonumber\\
\omega_{n, S^{5}}^{2} &=& n^{2} + 0\cdot q + \frac{n^{2}}{8(n^{2}-1)}\,(q^{2}+q^{3})+\dots,
\qquad \qquad |n|\neq 1 \ , \la{nenu}
\\
\omega_{1, S^{5}}^{2} &=& 1 + \frac{1}{2} q + \frac{11}{32}q^{2}
+ \frac{17}{64}q^{3}+\dots, \ \ \ \ \ \ \ \ \ \ \ \ \ \
\omega_{-1, S^{5}}^{2} = 1 - \frac{1}{2} q - \frac{5}{32}q^{2}
- \frac{5}{64}q^{3}+\dots, \no  \\
\omega_{n, F}^{2} &=& n^{2} + q\frac{2n^{2}}{4n^{2}-1} + q^{2}
\frac{n^{2}(-5-32n^{2}+80n^{4})}{4(-1+4n^{2})^{3}} \nonumber\\
&&+\  q^{3} \frac{n^2 \left[2 \left(960 n^6-816 n^4+60 n^2+79\right) n^2+3\right] }{8 \left(4 n^2-1\right)^5}
+\dots\quad . \nonumber
\ea
Treating separately the $n=0, -1, 1$ and $|n|\ge 2$ terms  in \rf{qqrr}  we find for the 2d energy
\begin{equation}
E_{2d} = 2\kappa + q (2-4\log 2)+q^{2}\Big(
\frac{5}{8}-\frac{5}{2}\log 2+\frac{3}{8}\zeta_{3}
\Big) +q^{3}\Big(
\frac{3}{8}-\frac{15}{8}\log 2+
\frac{45}{64}\zeta_{3}-\frac{15}{64}\zeta_{5}
\Big)+\dots.
\end{equation}
Dividing by $\kappa$ and expressing everything in terms of $\mathcal J_{2}$, we arrive at
\ba
E_{1} &=& 2+\sqrt{2\mathcal J_{2}}  \  \Big[2-4\log 2-
\mathcal J_{2}\Big(\frac{1}{2}+\frac{3}{2}\log 2-\frac{3}{4}\zeta_{3}
\Big)       \nonumber\\
&&\ \ \ \  \ \ \ \ \ \ \ \ \ \ \ \ \ \ \   +\ \mathcal J_{2}^{2}\Big(
\frac{1}{64}-\frac{15}{32}\log 2+\frac{51}{32}\zeta_{3}-\frac{15}{16}\zeta_{5}
\Big)+\dots\Big].  \la{chac}
\ea

\renewcommand{\theequation}{7.\arabic{equation}}
 \setcounter{equation}{0}
\setcounter{section}{6}

\section{Summary and concluding remarks}

In this paper we continued  the investigation  \cite{bd}
of the exact structure of  one-loop correction to energy of an  important class of classical string solutions
in \adss  expressed in terms
simple elliptic functions.  This  elliptic class  is next in complexity  to the simplest rational  class \ci{ft03,art}   for which the  classical
solutions are expressed in terms of linear or trigonometric  functions  of world-sheet coordinates
and thus the quadratic fluctuation operators can be  put into the form where their  coefficients are constant and
thus their  spectrum
can be easily  found.

The elliptic  solution considered  in  \cite{bd} was the folded spinning  string in $AdS_5$
for which it was shown that the  quadratic fluctuation operators can be put into  the standard single-gap L\' ame form;
 that allows one  to compute the corresponding determinants and thus the one-loop correction to the string energy
exactly for any value of semiclassical spin parameter $\S$.
Here we have demonstrated that the  same is true also for other  basic  elliptic solutions:
the pulsating string in $\mathbb R\times S^{2}$, the pulsating string in $AdS_3$ and  the folded spinning string in
 $\mathbb R\times S^{2}$.
 In all of  these cases where
there is only one charge/adiabatic invariant besides the energy,  namely, an  oscillator
 number or spin (in $S^5$ or $AdS_5$),
the  fluctuation operators  can be decoupled   and
 put into  a single-gap Lam\'e type form.
In fact, there is an explicit analytic continuation between the pulsating string and folded string cases. For example, in the $\mathbb R\times S^{2}$ case, the mapping $i\,\omega\leftrightarrow\sqrt{m^2-\kappa^2}$  maps the classical conserved quantities into one another: $\omega\, {\mathcal J}_2\leftrightarrow i\, {\mathcal N}$, as can be seen from (\ref{nan}) and (\ref{j2}) after using some elliptic function identities. Moreover,  the fluctuation operators also map into one another, with the identification:  $\omega\,\sigma \leftrightarrow \sqrt{m^2-\kappa^2}\,\tau +{\mathbb K}\left(\frac{\kappa^2}{\kappa^2-m^2}\right)$.
While this does not directly  imply the
   equivalence of the corresponding expressions for the
one-loop energies,  this relation is quite intriguing and is worth further study.

We have found, in particular, the expansion of the one-loop  energies
in the  limit of small values of the semiclassical parameters  corresponding to small size of the string.
This is equivalent  to the ``near-flat''  approximation    when  the string  probes
only small region of  \adss so that  its energy should start with   the  standard
 flat-space form plus corrections due to curvature.
As was argued in \ci{tt,rt} this  ``short-string''
 limit (in which the finite-size effects of the compact $\sigma\equiv  \sigma + 2 \pi$  string direction are all taken into account)
 may shed light on the structure of  strong-coupling corrections to dimensions of ``short''  dual  gauge
theory operators for which the  ``wrapping'' contributions are important.\foot{While the result of \ci{grom} guarantees
that the strong coupling expansion of TBA in similar semiclassical limit
should reproduce the full string semiclassical one-loop  correction,   that was shown only   in the  $sl(2)$ sector
in the limit  when the orbital momentum  $\J$   in $S^5$  is non-zero.
The limit $\J \to 0$ may be subtle, and therefore  the explicit string-theory  results  provide important data points.}

The semiclassical approximation is based on assumption  that $\sql \gg 1$ with
semiclassical parameters like $ \S= \frac{ S }{ \sql}, \ \J = \frac{ J }{\sql}$ or $\N= \frac { N }{ \sql}$
fixed, so that  $S, J$ or $N$ are   formally  large.
Still, taking  the  ``short-string''  limit  in which $\S, \J, \N \to 0$   one may
conjecture that if  that limit ``commutes''   with large the  $\sql$ limit it may shed  light on the form of
the quantum string energies with  fixed (e.g., small) values of  the spins and oscillation numbers $(S,J, N)$.
While this conjecture is hard to justify at the moment, the study of  the ``short-string'' limit
appears to provide some  qualitative  information on the structure of the large tension expansion  of quantum   string energies or
strong-coupling expansion of dimensions of dual
gauge-theory operators.

\subsection{Results}

Below we  summarize the results for the ``short-string'' (small spin or oscillation number)
 expansion of the classical
$E_0$  and one-loop $E_1$
energies of the four basic  elliptic \adss solutions analysed in  \cite{bd} and here:
folded  spinning strings  in   $\mathbb R\times S^{2}$   and  $AdS_3$,
and pulsating circular strings in  $\mathbb R\times S^{2}$   and  $AdS_3$.
We consider the case  of minimal winding number $m=1$.
We recall our notation:
$E=E_0 + E_1 + ...$, $E_0 = \sql  \mathcal E_{0} $,\   $E_1= \mathcal E_{1}$.
Also,  the non-zero  spin in $S^2$ is   $J_2\equiv J$.



\underline{Folded spinning string in $\mathbb R\times S^{2}$  }
\ba
\mathcal E_{0} &=& \sqrt{2\,\mathcal J}\,\Big(1+\frac{1}{8}\,\mathcal J
+\frac{3}{128}\mathcal J^{2}+\dots\Big), \no \\
   E_{1} &=& 2+\sqrt{2\,\mathcal J}\,\Big[2-4\,\log 2 + \Big(
-\frac{1}{2}-\frac{3}{2}\log 2+\frac{3}{4}\zeta_{3}
\Big)\,\mathcal J  \no  \\
 & &  \ \ \ \ \  +    \Big(\frac{1}{64}-\frac{15}{32}\log 2+\frac{51}{32}\zeta_{3}-\frac{15}{16}\zeta_{5}\Big)\,\mathcal J^{2}
+\dots
\Big], \nonumber \\
E &=& \sqrt{2J  \sqrt\lambda}\Big(1+\frac{\frac{1}{8}J+2-4\log 2}{\sqrt\lambda}+\dots\Big)+2+\dots
\la{ro1}
\ea

\underline{Folded  spinning  string in $AdS_{3}$}
\ba
\mathcal E_{0} &=& \sqrt{2\,\mathcal S}\,\Big(1+\frac{3}{8}\,\mathcal S-\frac{21}{128}\mathcal S^{2}
+\dots\Big),\no  \\
   E_{1} &=& 1+\sqrt{2\,\mathcal S}\,\Big[\frac{3}{2}-4\,\log 2 + \Big(
-\frac{23}{16}+\frac{3}{2}\log 2+\frac{3}{4}\zeta_{3}
\Big)\,\mathcal S    \no \\
 & & \ \ \ \ \ \   +
\Big(
\frac{689}{256}-\frac{63}{32}\log 2-\frac{15}{32}\zeta_{3}-\frac{15}{16}\zeta_{5}
\Big)\,\mathcal S^{2}+\dots
\Big], \nonumber \\
E &=& \sqrt{2S\sqrt\lambda}\Big(1+\frac{\frac{3}{8}S+\frac{3}{2}-4\log 2}{\sqrt\lambda}+\dots\Big)+1+\dots
\la{ro2}
\ea

\underline{Pulsating string in $\mathbb R\times S^{2}$}
\ba
\mathcal E_{0} &=& \sqrt{2\,\mathcal N}\,\Big(1-\frac{1}{8}\,\mathcal N-\frac{5}{128}\,\mathcal N^{2}+\dots\Big),\no  \\
  E_{1} &=& 2+\sqrt{2\,\mathcal N}\,\Big[1-4\,\log 2 + \Big(
\frac{1}{8}+\frac{3}{2}\log 2+\frac{3}{4}\zeta_{3}
\Big)\,\mathcal N    \no \\
 & & \ \ \ \ \    +   \Big(
\frac{11}{128} + \frac{25}{32}\log 2-\frac{135}{32}\zeta_{3}-\frac{15}{16}\zeta_{5}
\Big)\mathcal N^{2}+\dots
\Big],\nonumber \\
E &=& \sqrt{2N\sqrt\lambda}\Big(1+\frac{-\frac{1}{8}N+1-4\log 2}{\sqrt\lambda}+\dots\Big)+2+\dots\,
\la{pu1}
\ea

\underline{Pulsating string in $AdS_{3}$}
\ba
\mathcal E_{0} &=& \sqrt{2\,\mathcal N}\,\Big(1+\frac{5}{8}\,\mathcal N-\frac{77}{128}\,\mathcal N^{2}+\dots\Big),\no
 \\
  E_{1} &=& 1+\sqrt{2\,\mathcal N}\,\Big[
\frac{5}{2}-4\log 2+\Big(
-\frac{37}{8}+\frac{5}{2}\log 2+\frac{3}{4}\zeta_{3}
\Big)\,\mathcal N    \no \\
 & & \ \ \ \ \   +
\Big(
\frac{3915}{256}-\frac{231}{32}\log 2-\frac{117}{32}\zeta_{3}-\frac{15}{16}\zeta_{5}\Big)\,\mathcal N^{2}
+\dots
\Big],\nonumber \\
E &=& \sqrt{2N\sqrt\lambda}\Big(1+\frac{\frac{5}{8}N+\frac{5}{2}-4\log 2}{\sqrt\lambda}+\dots\Big)+1+\dots,
\la{pu2}
\ea
We observe a remarkable universality of the  small charge expansion of the energy of
 all four elliptic solutions.\foot{One may wonder if the  1-loop expressions we found are scheme-dependent.
 The choice of scheme preserving all relevant   symmetries
  in computations with GS action is a subtle issue that deserves further study
 (see \ci{rtt07} for a discussion).
 Here as in several  previous papers we assumed that bosonic and fermionic  contributions are first
 added  together and then the (finite) sum over modes is performed.}
  In particular, the  leading terms with  transcendental  coefficients
($\log 2, \ \zeta_{3}, \ \zeta_{5}, ...$)  happen to have the same form.
Compared to  similar expansions for rational  rigid spinning  string solutions
discussed in \ci{rt}   we notice  the presence of the $\log 2$ term  already in the leading one-loop coefficient
which   was absent in the rational case. \foot{The universality of the $\log 2$ coefficient
  suggests that maybe it can be
absorbed into a redefinition of $\l$ (cf. cusp anomaly case \ci{bkk}).
Indeed, a simple shift of $\sql$ by $4 \ln 2$ removes the leading $\ln 2$ terms in the 1-loop correction,
but it does not remove $\ln 2$ coefficients  in subleading terms so we are not sure if
that this  shift  may have  a deeper meaning.}

\subsection{Interpolation to finite quantum numbers:\\
energies of strings corresponding to  first excited string level }

The semiclassical approximation   discussed above  was based on assumption that
one takes $\sql \gg 1$ for fixed $ \mathcal Q = { Q \ov \sql }$
with $Q= (N, S, J, ...)$ and then expands in $ \mathcal Q \to 0$.
This  still means that $Q \gg 1$.
As was argued  in \ci{rt}, if one assumes that the resulting expressions for string energy
can be formally interpolated  to finite values of $Q$
they should then describe leading corrections to energies of the corresponding
quantum string states. In particular,
one may consider the  analogs of  states at the first excited
string
level  which should correspond to   members of the Konishi
multiplet \ci{bian,rt}  (if this is the case   their energies should differ
only by $\l$-independent half-integer constants).

Interpolation  from semiclassical expressions for $E$ like given  above, i.e.
$E = \sqrt{ 2Q \sql } \big(1 + { a Q + b \ov \sql } + ...\big)  $  valid for $\sql \gg 1$ and
fixed $ \mathcal Q \ll 1$, i.e.    $ Q\gg 1$,
to quantum string energies  with finite $Q$ is, of course, potentially ambiguous.
One requirement is that one should match  the corresponding flat-space expressions.
In \ci{rt} this ambiguity was fixed by shifting $Q \to Q-2$  everywhere in the expression for the energy, i.e.
$E(\sql, Q)  \to   E(\sql, Q-2)$.
An alternative recipe  that we shall consider here is   to  do this  shift only
under the square root
\be \la{sta}
E = \sqrt{ 2(Q-2) \sql }\ \Big(1 + { a Q + b \ov \sql } + ...\Big)  \ . \ee
One may think that this is  suggested  by the structure of the
  solution of the  the marginality condition for the
corresponding vertex operator  which looks like  $ 2=Q  - \frac{ 1 }{ 2 \sql}
 [ E(E -4) - a Q (Q+ b)+ ...  ] + ...$,
see \ci{rt}.\foot{At the same time, the  recipe of \ci{rt}, i.e. $Q\to Q-2$,  may be motivated by
the requirement that not only the leading term but also $Q$-dependent
corrections should vanish for the  BPS  ground-state cases with  $Q=2$.}
Then to get    the  energies of states on the first excited string level we should
start with \rf{sta} and
set  $Q=(N,J,S)=4$
(for  states on the first excited string level  the corresponding vertex operators should
contain factors like
$(\del x \bar \del x )^{Q/2} = (\del x  \bar \del x )^2$, etc.).

Using the above
results 
 \rf{ro1}--\rf{pu2}
 we then get
\ba
&& E_{\rm folded\      \mathbb R\times S^{2}} = 2\fl \Big(1+\frac{\frac{5}{2}-4\log 2}{\sqrt\lambda}+\dots\Big)+2 \ , \la{kh}
\\
&& E_{\rm folded\     AdS_3 }  =
 2\fl \Big(1+\frac{3-4\log 2}{\sqrt\lambda}+\dots\Big)+1
\  , \la{hkh}
\\
&& E_{\rm  pulsating \     \mathbb R\times S^{2} }  = 2\fl \Big(1+\frac{\frac{1}{2}-
4\log 2}{\sqrt\lambda}+\dots\Big)+2
\  , \la{pkh}
\\
&& E_{\rm pulsating \     AdS_3 }  = 2\fl \Big(1+\frac{5-4\log 2}{\sqrt\lambda}+\dots\Big)+1
\  . \la{skh}
\ea
The difference    of the  coefficients of the first subleading term here
and for the rational  solutions in    \cite{rt}
 may be due to the fact that these states are not actually  in the
same supermultiplet  so  dimensions need  not  be related  just by an integer
 number shift.\foot{For example, the
folded string in $AdS_5$  without  orbital momentum in $S^5$
  may be dual to  an
operator  built out field strengths like  Tr$( F D^S F)$
 that mixes with other  similar operators
and is not in the  Konishi multiplet.
In general, the question of identification of states  in semiclassical expansion
is subtle as finite values of $S^5$ orbital momentum  cannot be resolved, so
one  cannot a priori distinguish between a  state dual to
 Tr$( F D^S F)$  and a state dual to Tr$( \Phi D^S \Phi)$.}
An alternative is that the semiclassical expressions cannot be actually interpolated to
 fixed values of quantum numbers.
   That issue  remains to be clarified; still,
  the similarity of the above expressions
and those in \ci{rt} for  energies of ``small'' strings suggest  that they
do model  quantum string energies, i.e.  are not very much off the mark.

\

\

{\bf Note added}

In discussing one-loop corrections for pulsating strings in section 5 we have tacitly
assumed that the fermions in \rf{fef} or \rf{A15}  with the  angular (``polar'') choice of
global coordinates like in \rf{A16}
 are periodic  in $\s$ for any value of the winding number $m$.
 As was   pointed out to us  by  Victor Mikhaylov
 after the first version of this paper appeared on the arXiv,
 this may  be  unnatural in view of the discussion in \cite{Mikhaylov}:
the fermions should be periodic for any $m$  in ``cartesian''   coordinates  but
 that  implies that
they should be antiperiodic for $m$=odd in ``polar'' coordinates  \ci{mikh}.
For example, in flat space, changing coordinates from  cartesian to polar  seems to require
rotation of the GS fermions  $\theta$ (target-space spinors)
 by an angle $\phi$, so that for a circular  solution
with $\p= m \sigma$  starting with $\theta(\s+2 \pi) = \theta(\s)$
one should end up with $\tilde \theta(\s+2 \pi) = (-1)^m\tilde
 \theta (\s)$.\foot{As for the folded string  in $AdS_3$ or
  $\mathbb R\times S^{2}$,
there is  no  obvious reason  to change from periodic to
  antiperiodic fermions (the corresponding rotation from cartesian to polar coordinates
  is $\tau$-dependent).
 Let us note, however,  a somewhat special ``singular'' nature of the folded string
 which may be considered
   as a special case of a spiky string \ci{spike}  which does encircle the origin like a
   pulsating string.
   }
We do not, however, find this reasoning convincing  since in curved space (or in general
coordinates)  the target-space spinors  do not transform  under diffeomorphisms
but rotate  under  local  Lorentz frame transformations (with the tangent-space metric
and  Dirac $\Gamma$-matrices being the standard  Minkowski ones
  for any choice of     the
coordinate labels).\foot{It is true, of course, that there is only one antiperiodic
spin structure on the disc  which is  familiar in the open  NSR string case
but in the  closed  string  case where the image  of the world sheet in the target
space should
be a cylinder the situation is different. For example, in the static gauge where
$t=\kappa \tau, \ \phi= m \sigma$   become arguments  of $\theta$
the latter  should still be defined  on a cylinder.}
To clarify this further,  in Appendix F
we discuss the fermionic kinetic term in the light-cone gauge  adding also
angular momentum in $S^5$ that allows one  to interpolate to the BMN limit. In the case of
pulsating string in flat space we explicitly show that the issue of periodicity/antiperiodicity
of fermions is indeed a gauge/coordinate artifact.


\


\section*{Acknowledgments }
We thank V. Forini, N. Gromov,  M. Kruczenski, V. Mikhaylov, R. Roiban  and
B. Vicedo for many useful discussions. G.D. acknowledges the DOE  grant DE-FG02-92ER40716, and
 A.T. acknowledges the support of the  Purdue University.

 \bigskip


\section*{Appendices}


\renewcommand{\theequation}{A.\arabic{equation}}
\renewcommand{\thesection}{A}
 \setcounter{equation}{0}
\setcounter{section}{1} \setcounter{subsection}{0}

\section{Fluctuation Lagrangian for  pulsating solution in $\mathbb{R} \times S^2$}
\label{appA}

Here we present details of computation of  fluctuation Lagrangian
for pulsating string  in $\mathbb{R} \times S^2$.

In conformal gauge
the $AdS_5$ part of fluctuation Lagrangian contains one massless mode (fluctuation of $t$)
and 4 massive modes with mass $\k$.
The $S^5$ part of  Lagrangian  written in terms of complex combinations of 6 embedding coordinates is
\be
L_{S} = -\frac{1}{2}\, \partial_a Z_i\,\partial^a Z_i^* + \frac{\Lambda}{2}(Z_i Z_i^*-1),
\ee
where for  pulsating solution
\be
Z_1 = \cos\psi(\tau),\qquad
Z_2 = \sin\psi(\tau)\,e^{i\,m\,\sigma},\qquad Z_3 = 0, \qquad \Lambda  = 2\, m^{2}\,\sin^{2}\psi-\kappa^{2}.
\ee
The  fluctuations of $Z_i$  satisfying
$
Z_i\,\widetilde Z_i^* + Z_i^*\,\widetilde Z_i = 0$
contain two massive modes $\widetilde Z_3$  with mass
$
M_3^2 = -\Lambda = \kappa^{2}-2\,m^{2}\,\sin^{2}\psi
$
and two coupled  fluctuations
\be
Z_1 = \cos\psi+g_1+i\,z,\qquad
Z_2 = (\sin\psi+g_2+i\,\xi)\,e^{i\,m\,\sigma}, \ \ \ \ \ \ \ \
\zeta \equiv  g_1\,\cos\psi + g_2\,\sin\psi =0 \ .
\ee
Introducing $\eta = g_2\,\cos\psi  - g_1\,\sin\psi$ orthogonal to $\zeta$ we end up with
the following fluctuation Lagrangian for the 3 remaining modes\footnote{We thank I. Park for pointing out to us some typos in this Appendix.}
\ba
\td L &=& \phantom{+}\frac{1}{2}(\dot \eta^2-\eta'^2-M_\eta^2\,\eta^2)
 + \frac{1}{2}(\dot z^2-z'^2-M^2\,z^2)
+ \frac{1}{2}(\dot \xi^2-\xi'^2-M_{\xi}^2\,\xi^2) + \nn
&&\ \ \ \ \ \ \ \ \ +\ m\,\cos\psi\,(\xi \,\eta'-\xi'\,\eta).
\ea
To decouple $\eta$ and $\xi$  fluctuations we may use the linearized Virasoro
constraints (see \ci{bd} for a similar discussion)
\ba
 m\,\sin\psi\left(m\,\cos\psi\,\eta+\xi'\right)-\kappa\,\dot\beta+\dot\psi\,\dot\eta = 0, \  \ \ \ \
 m\sin\psi\dot \xi_{2}-m\,\cos\psi\,\dot\psi\,\xi-\kappa\beta'+\dot\psi\eta'=0 \ ,
\ea
where $\beta$ is the massless mode from $AdS_5$.
Using the equations of motion  for $\eta$ and $\xi$ fluctuations
written for   $\sim e^{i\,n\,\sigma}$ Fourier mode in $\s$
\ba
-\ddot \eta-(n^{2}+M_{\eta}^{2})\eta-2\,i\,m\,n\,\cos\psi\,\xi &=& 0, \\
-\ddot \xi-(n^{2}+M_{\xi}^{2}) \xi+2\,i\,m\,n\,\cos\psi\,\eta &=& 0.
\ea
we get
$
\eta = \frac{\ddot \xi+(n^{2}+M_{\xi}^{2})\xi}{2\,i\,m\,n\,\cos\psi},
$
and  thus  obtain the following equation for $\xi$
\be
\frac{1}{\cos\psi}(\partial_{\tau}^{2}+n^{2}+M_{\eta}^{2})
\frac{1}{\cos\psi}(\partial_{\tau}^{2}+n^{2}+M_{\xi}^{2})\,\xi-4\,m^{2}\,n^{2}\,\xi=0. \label{jros}
\ee
This equation can be written in a  factorized form
\be
\frac{1}{\sin\psi\cos\psi}(\partial_{\tau}^{2}+\mu^{2})\frac{\sin^{2}\psi}{\cos\psi}(\partial_{\tau}^{2}+n^{2})
\frac{\xi}{\sin\psi} = 0,
\ \ \ \ \ \ \ \ \ \
\mu^{2} = n^{2} +\kappa^{2}\Big(1-\frac{2}{\sin^{2}\psi}\Big).
\ee
Thus we end up  with two decoupled modes --  a massless mode and a mode with mass
 $M^2= \kappa^{2} \big(1-\frac{2}{\sin^{2}\psi}\big)$.

The same decoupling  happens  directly if we start with the Nambu action  and use  the static gauge on the fluctuations
of  $t$ and $\phi$.
If we parametrise the metric as in \cite{rtt07}  ($k=1,2,3,4$)
\begin{eqnarray}
ds^2 &=&-\Big(\frac{1+\frac{1}{4}\eta^2}{1-\frac{1}{4}\eta^2}\Big)^2 d t^2+ \frac{d\eta_k d \eta_k }{(1-\frac{1}{4}\eta^2)^2}\nonumber\\
&+& \frac{dx^2 +dy^2 -(x dy-y dx)^2}{1-x^2-y^2}+(1-x^2-y^2)( d \psi^2 + \cos^2 \psi\  d \varphi^2 +\sin^2 \psi \ d \phi^2)
\end{eqnarray}
so that pulsating solution is
\begin{equation}
t=\kappa \tau, \quad \quad \eta_k=0, \quad \quad x=y=0, \quad \quad \psi=\psi(\tau), \quad \quad \varphi=0, \quad \quad \phi=m \sigma
\end{equation}
Expanding the Nambu action with  $\tilde t=0$ and $\tilde \phi=0$
we get
\begin{eqnarray}
\tilde{S}&=&\frac{\sqrt{\lambda}}{4 \pi}\int d \tau d \sigma \bigg[(\partial_{\sigma} \tilde{\eta}_k)^2-(\partial_{\tau} \tilde{\eta}_k)^2+ \kappa^2 \tilde{\eta}_k^2+ (\partial_{\sigma} \tilde{x})^2-(\partial_{\tau} \tilde{x})^2+ (\kappa^2-2 m^2 \sin^2 \psi) \tilde{x}^2\nonumber\\
&+&(\partial_{\sigma} \tilde{y})^2-(\partial_{\tau} \tilde{y})^2+ (\kappa^2-2 m^2 \sin^2 \psi) \tilde{y}^2 +
\cos^2 \psi [ (\partial_{\sigma} \tilde{\varphi})^2-(\partial_{\tau} \tilde{\varphi})^2]\nonumber\\
&+& \frac{\kappa^2}{m^2}\frac{1}{\sin^2 \psi}[ (\partial_{\sigma} \tilde{\psi})^2-(\partial_{\tau} \tilde{\psi})^2]-m^2 \sin^2 \psi \tilde{\psi}^2 - \partial_{\tau} \psi \tilde{\psi} \partial_{\tau} \tilde{\psi}  \cot \psi \bigg].
\end{eqnarray}
With the field redefinitions
\begin{equation}
\cos \psi\ \tilde{\varphi}= \xi, \quad \quad \frac{\kappa}{m}\frac{1}{\sin \psi}\tilde{\psi}=g
\end{equation}
we finish with
\begin{eqnarray}
\tilde{S}&=&\frac{\sqrt{\lambda}}{4 \pi}\int d \tau d \sigma \bigg[(\partial_{\sigma} \tilde{\eta}_k)^2-(\partial_{\tau} \tilde{\eta}_k)^2+ \kappa^2 \tilde{\eta}_k^2+ (\partial_{\sigma} \tilde{x})^2-(\partial_{\tau} \tilde{x})^2+ (\kappa^2-2 m^2 \sin^2 \psi) \tilde{x}^2\nonumber\\
&+&(\partial_{\sigma} \tilde{y})^2-(\partial_{\tau} \tilde{y})^2+ (\kappa^2-2 m^2 \sin^2 \psi) \tilde{y}^2 +
(\partial_{\sigma} \xi)^2-(\partial_{\tau} \xi)^2+ (\kappa^2-2 m^2 \sin^2 \psi) \xi^2\nonumber\\
&+& (\partial_{\sigma} g)^2-(\partial_{\tau} g)^2 + \kappa^2 (1-\frac{2}{\sin^2 \psi}) g^2 \bigg]
\end{eqnarray}
To get the fermionic  part of the fluctuation Lagrangian
we start  with the standard form of the action (see, e.g., \ci{dgt,ft02})
\be\la{A15}
{\cal L}_F = -2\,i\,\overline\vartheta\left(-\rho^a\,D_a-\frac{i}{2}\varepsilon^{ab}\,\rho_a\,\Gamma_*\,\rho_b\right)\,\vartheta.
\ee
Using  the standard choice of global angular \adss coordinates  we  have
\be\la{A16}
\begin{array}{c|cccccccccc}
\mu & 0 & 1 & 2 & 3 & 4 & 5 & 6 & 7 & 8 & 9 \\
\hline
X^\mu & \kappa\tau & 0 & 0 & 0 & 0 & \pi/2 & 0 & \psi(\tau) & 0 & m\sigma
\end{array}
\ee
The non zero vielbein  and spin connection components are
\ba
&&E^{0}_{0} = E^{1}_{1} = E^{5}_{5} = E^{7}_{7} = 1, \qquad E^{8}_{8} = \cos\psi, \qquad
E^{9}_{9} = \sin\psi, \\
&&\omega^{12}_{2} = -1, \quad  \omega^{13}_{3} = -1, \quad \omega^{24}_{4} = -1, \quad
\omega^{56}_{6} = 1, \quad \omega^{78}_{8} = \sin\psi, \quad \omega^{79}_{9} = -\cos\psi,
\ea
so that
\ba
D_{\tau} = \partial_{\tau}, \qquad
D_{\sigma} = \partial_{\sigma} -\frac{1}{2} m \cos\psi \,\Gamma_{79} , \ \ \ \ \ \
\rho_{\tau} = \kappa\,\Gamma_{0}+\dot\psi\, \Gamma_{7}, \qquad
\rho_{\sigma} = m\sin\psi\,\Gamma_{9}
\ea
and finally
$
{\cal L}_F = -2\,i\,\overline\vartheta\,D_F\,\vartheta,
$
with
\ba
D_F = (\kappa\Gamma_{0}+\dot\psi\Gamma_{7})\partial_{\tau}-m\,\sin\psi\,\Gamma_{9}\,\partial_{\sigma}
 -\frac{m^{2}}{2}\,\sin\psi\cos\psi\,\Gamma_{7}+m\,\sin\psi\,\dot\psi\Gamma_{07}
\,\Gamma_{12349}
\ea
Performing  a Lorentz rotation
\be
\Gamma_0(s) = e^{\frac{1}{2} s \Gamma_{07}}\Gamma_0 e^{-\frac{1}{2} s \Gamma_{07}},  \ \ \ \ \ \ \ \ \  \
\Gamma_7(s) = e^{\frac{1}{2} s \Gamma_{07}}\Gamma_7 e^{-\frac{1}{2} s \Gamma_{07}}.
\ee
with
$
\sinh s = -\frac{\dot\psi}{\sqrt{\kappa^2-\dot\psi^2}},\ \ \cosh s = \frac{\kappa}{\sqrt{\kappa^2-\dot\psi^2}},
$
we obtain
$$
D_F' = e^{\frac{1}{2} s \Gamma_{07}} D_F e^{-\frac{1}{2} s \Gamma_{07}} =
m\sin\psi\Big[\Gamma_{0}(\partial_{\tau}-\frac{1}{2}\Gamma_{07}\partial_{\tau}s)-\Gamma_{9}\,\partial_{\sigma}\Big]
-\frac{m}{2}\,\cos\psi\,(\kappa\Gamma_{7}-\dot\psi\Gamma_{0})
 +m\,\sin\psi\,\dot\psi\Gamma_{07}
\,\Gamma_{12349}
$$
Simplifying this  we get
\ba
D_F'  =
m\sin\psi\left(\Gamma_{0}\partial_{\tau}-\Gamma_{9}\,\partial_{\sigma}\right)
+\frac{m}{2}\,\dot\psi\cos\psi\,\Gamma_{0}
 +m\,\sin\psi\,\dot\psi\Gamma_{07}
\,\Gamma_{12349}
\ea
Rescaling of the fermions
$
\vartheta\to \vartheta\frac{1}{\sqrt{m\sin\psi}}, \qquad D_{F}'' = \frac{1}{\sqrt{m\sin\psi}}\,D_F' \,\frac{1}{\sqrt{m\sin\psi}},
$
gives
\be
D_{F}'' =
\Gamma_{0}\partial_{\tau}-\Gamma_{9}\,\partial_{\sigma}+\dot\psi\,\Gamma_{0123479}.
\ee
Diagonalizing $\Gamma_{1234} = \pm 1$ we end with
\be
D_{F} = \Gamma_{0}\partial_{\tau}-\Gamma_{9}\,\partial_{\sigma}+\dot\psi\Gamma_{079}
\ee


\renewcommand{\theequation}{B.\arabic{equation}}
\renewcommand{\thesection}{B}
 \setcounter{equation}{0}
\setcounter{section}{1} \setcounter{subsection}{0}

\section{Fluctuation Lagrangian for  pulsating solution in $AdS_3$}
\label{appB}


Here we shall use  the following coordinates
\begin{eqnarray}
ds^2 &=& - \cosh^2 \rho dt^2 + d \rho^2 + \sinh^2 \rho [d \beta_1^2 +\cos^2 \beta_1 (d \beta_2^2 +\cos^2 \beta_2 d \beta_3^2)]\nonumber\\
&+& d \psi_1^2 +\cos^2 \psi_1[d \psi_2^2 + \cos^2 \psi_2 (d \psi_3^2 +\cos^2 \psi_3 (d \psi_4^2 + \cos^2 \psi_4 d \psi_5^2))]
\end{eqnarray}
The pulsating solution in these coordinates is
\begin{equation}
\beta_1=\beta_3=0, \quad \quad \beta_2=m \sigma, \quad t=t(\tau), \quad \rho=\rho(\tau), \quad \psi_i=0
\end{equation}
Fixing the fluctuations of $t$ and $\beta_2$ to zero and  expanding  the Nambu-Goto action we
 obtain the following fluctuations Lagrangian for the physical  8 fields
\begin{eqnarray}
\tilde{L}&=&\frac{1}{2} \Big[\sinh^2 \rho [(\partial_{\sigma} \tilde{\beta}_1)^2- (\partial_{\tau} \tilde{\beta}_1^2)-m^2 \tilde{\beta}_1^2]+ \sinh^2 \rho \cos^2 m \sigma [(\partial_{\sigma} \tilde{\beta}_3)^2- (\partial_{\tau} \tilde{\beta}_3^2)]\nonumber\\
&+& \frac{4\k^2}{m^2}\frac{1}{\sinh^2 (2 \rho)}[(\partial_{\sigma} \tilde{\rho})^2-(\partial_{\tau} \tilde{\rho})^2]+
[ \frac{\k^2(1+ 2 \cosh 2 \rho)}{ \cosh^4 \rho}-\frac{\k^4}{ m^2 \cosh^6 \rho}+m^2 \sinh^2 \rho]\tilde{\rho}^2\nonumber\\
&+& \frac{8 \k^2 -3 m^2 -m^2(4 \cosh 2 \rho +\cosh 4 \rho)}{4 m^2 \sinh \rho \cosh^3 \rho}
 \ \partial_{\tau} \rho \ \tilde{\rho} \ \partial_{\tau}  \tilde{\rho}  +(\partial_{\sigma} \tilde{\psi}_i)^2-(\partial_{\tau} \tilde{\psi}_i)^2\Big]
\end{eqnarray}
where $\k=\E_0$ is the integration constant in \rf{glo}.
After  the field redefinitions
\begin{equation}
\tilde{\beta}_3 \cos m \sigma \sinh \rho  = \eta, \quad \quad \tilde{\beta}_1 \sinh \rho= \xi, \quad \quad \frac{2 \k}{m \sinh 2 \rho} \tilde{\rho}= \zeta
\end{equation}
the fluctuation Lagrangian becomes (after integration by parts)
\begin{eqnarray}
\tilde{L}&=&\frac{1}{2}\bigg[ (\partial_{\sigma} \tilde{\psi}_i)^2-(\partial_{\tau} \tilde{\psi}_i)^2 + (\partial_{\sigma} \eta)^2-(\partial_{\tau} \eta)^2+ 2 m^2 \eta^2 \sinh^2 \rho +(\partial_{\sigma} \xi)^2-(\partial_{\tau} \xi)^2+ 2 m^2 \xi^2 \sinh^2 \rho \nonumber\\
&+& (\partial_{\sigma} \zeta)^2-(\partial_{\tau} \zeta)^2+ \zeta^2 (2 m^2 \sinh^2 \rho - \frac{2 \k^2}{\sinh^2 \rho}) \bigg]
\end{eqnarray}
To find the fermionic Lagrangian we label directions as
\be
\begin{array}{c|ccccc}
\mu & 0 & 1 & 2 & 3 & 4  \\
\hline
X^\mu & t(\tau) & \rho(\tau) & 0  &  m\sigma  & 0
\end{array}
\ee
The relevant non zero vielbein and connection  components are
\ba
E^{0}_{0} = \cosh\rho, \quad E^{1}_{1} = 1, \quad E^{2}_{2} = \sinh\rho,
\quad E^{3}_{3} = \sinh\rho \ , \\
\omega^{01}_{0} = \sinh\rho, \quad \omega^{12}_{2}=-\cosh\rho, \quad
\omega^{13}_{3} = -\cosh\rho, \quad \omega^{24}_{4} = -1,
\ea
so that
\ba
D_{\tau} &=& \partial_{\tau}+\frac{1}{2}\,\k\,\frac{\sinh\rho}{\cosh^{2}\rho}\,\Gamma_{01}, \quad
D_{\sigma} = \partial_{\sigma} -\frac{m}{2}  \,\cosh\rho \,\Gamma_{13} \\
\rho_{\tau} &=& \cosh\rho\,\dot t\,\Gamma_{0} +\dot\rho\,\Gamma_{1} = \frac{\k}{\cosh\rho}
\,\Gamma_{0} +\dot\rho\,\Gamma_{1}, \quad
\rho_{\sigma} = m\,\sinh\rho\,\Gamma_{3}.
\ea
The fermionic  operator is then
\ba
D_{F} &=& -\rho^{a}D_{a}-i\,\rho_{\tau}\Gamma_{*}\rho_{\sigma} =  -\rho^{a}D_{a}+\rho_{\tau}\rho_{\sigma}\Gamma_{01234}  \no \\
&=& (\dot t\,\cosh\rho\,\Gamma_{0}+\dot\rho\,\Gamma_{1})\partial_{\tau}-m\sinh\rho\,\Gamma_{3}\partial_{\sigma}
-\frac{m^{2}}{2}\sinh\rho\cosh\rho\,\Gamma_{1} \\
&& +m\,\sinh\rho\,(\dot t\,\cosh\rho\,\Gamma_{0}+\dot\rho\,\Gamma_{1})\Gamma_{3}\Gamma_{01234}
+\frac{1}{2}\dot t\,\sinh\rho\,(\dot t\,\cosh\rho\,\Gamma_{0}+\dot\rho\,\Gamma_{1})\,\Gamma_{01}.\nonumber
\ea
Again, it is useful to perform a Lorentz rotation
\ba
\Gamma_0(s) = \Gamma_0\,\cosh s+\Gamma_1\,\sinh s, \qquad
\Gamma_1(s) = \Gamma_1\,\cosh s+\Gamma_0\,\sinh s.
\ea
with $
\sinh s = -\frac{\dot\rho}{\sqrt{\dot t^{2}\cosh^{2}\rho-\dot\rho^{2}}},\ \
\cosh s = \frac{\dot t\,\cosh\rho}{\sqrt{\dot t^{2}\cosh^{2}\rho-\dot\rho^{2}}},
$
Finally we get
\ba
D'_{F}
&=& m\,\sinh\rho\,(\Gamma_{0}\partial_{\tau}-\Gamma_{3}\partial_{\sigma}) + \frac{m}{2}\,\dot \rho\,\cosh\rho\,\Gamma_{0}+ m^{2}\sinh^{2}\rho\,\Gamma_{124}.
\ea
Rescaling the fermions by $\frac{1}{\sqrt{m\sinh\rho}}$ we end up with
\be
D_{F}= \Gamma_{0}\partial_{\tau}-\Gamma_{3}\partial_{\sigma}+m\,\sinh\rho\,\Gamma_{124}.
\ee

\renewcommand{\theequation}{C.\arabic{equation}}
\renewcommand{\thesection}{C}
 \setcounter{equation}{0}
\setcounter{section}{1} \setcounter{subsection}{0}

\section{Fluctuation Lagrangian for  folded spinning string   in $ \mathbb{R} \times S^2    $}
\label{appC}


Starting with the metric
\begin{eqnarray}
ds^2 &=& -\Big(\frac{1+\frac{1}{4}\eta^2}{1-\frac{1}{4}\eta^2}\Big)^2 d t^2+ \frac{d\eta_k d \eta_k }{(1-\frac{1}{4}\eta^2)^2}\nonumber\\
&+& d \psi_1^2 +\cos^2 \psi_1[d \psi_2^2 + \cos^2 \psi_2 (d \psi_3^2 +\cos^2 \psi_3 (d \psi_4^2 + \sin^2 \psi_4 d \phi^2))]
\end{eqnarray}
the folded spinning string  on $S^2$ of $S^5$ is
\begin{equation}
t=\kappa \tau, \quad \quad \eta_k=0, \quad \quad \psi_i=0, \quad \quad \psi_4= \theta(\sigma), \quad \quad \phi=w \tau
\end{equation}
where $k=1,2,3,4$  and  $i=1,2,3$.
Fixing the static gauge on  fluctuations by setting
 $\td t$ and $\td \psi_4$ to zero and expanding the Nambu action  we obtain
\begin{eqnarray}
\tilde{S}&=&\frac{\sqrt{\lambda}}{4 \pi}\int d \tau d \sigma \Big[(\partial_{\sigma} \tilde{\eta}_k)^2-(\partial_{\tau} \tilde{\eta}_k)^2+ \kappa^2 \tilde{\eta}_k^2+ (\partial_{\sigma} \tilde{\psi}_i)^2-(\partial_{\tau} \tilde{\psi}_i)^2+ (2 w^2 \sin^2 \theta -\kappa^2) \tilde{\psi}_i^2 \nonumber\\
&+&\frac{\kappa^2 \sin^2 \theta}{\theta'^2} [(\partial_{\sigma} \tilde{\phi})^2-(\partial_{\tau} \tilde{\phi})^2]\Big]
\end{eqnarray}
Setting
$
f= \frac{\kappa \sin \theta}{\theta'}\tilde{\phi}
$
we finally obtain
\begin{eqnarray}
\tilde{S}&=&\frac{\sqrt{\lambda}}{4 \pi}\int d \tau d \sigma \Big[(\partial_{\sigma} \tilde{\eta}_k)^2-(\partial_{\tau} \tilde{\eta}_k)^2+ \kappa^2 \tilde{\eta}_k^2+ (\partial_{\sigma} \tilde{\psi}_i)^2-(\partial_{\tau} \tilde{\psi}_i)^2+ (2 w^2 \sin^2 \theta -\kappa^2) \tilde{\psi}_i^2 \nonumber\\
&+& (\partial_{\sigma} f)^2-(\partial_{\tau} f)^2  + f^2 \kappa^2 (1- \frac{2 (\kappa^2-w^2)}{\theta'^2})\Big]
\end{eqnarray}
To find the  fermionic  Lagrangian we start with
$
{\cal L}_F = -2\,i\,\overline\vartheta\,D_F\,\vartheta.
$   where
 $\vartheta$ is  the  Majorana-Weyl 10d spinor,
$
\overline\vartheta = \vartheta^t\,\Gamma_0,\ \ \Gamma_{11}\vartheta=\vartheta
$
and we shall use real Gamma matrices (as in, e.g.,   ~\cite{Callan:2004uv}).
For more general 2-spin solution on $S^3$  we find
\ba
D_F &=& s_1\,\Gamma_0\partial_0-\Gamma_7\partial_1+u\,\Gamma_{078}\Gamma_{1234}+s_1\,\frac{\kappa w_1 w_2}{2u^2}\,\Gamma_{789}, \\
s_1 &=& {\rm sign}(\theta'), \ \ \ \ \ \ \ \ \ \
u = \sqrt{w_1^2\,\cos^2\theta + w_2^2\,\sin^2\theta}.
\ea
In the case of  $w_1=0, \ \ w_2=w\not=0$  this is
\be
D_F = s_1\,\Gamma_0\partial_0-\Gamma_7\partial_1+u\,\Gamma_{078}\Gamma_{1234}.
\ee
The functional integral over the   Majorana fermions gives the square root of the determinant
of the operator $\Gamma_0\,D_F$,
\be
\mathscr{D}_f = {\det}^{1/2}_{\rm Weyl}\, (\Gamma_0\,D_F).
\ee
The operator  $\Gamma_0\,D_F$ has a block structure respecting the Weyl condition
$
[\Gamma_0 D_F, \Gamma_{11}]=0.
$
 The spectrum is Weyl symmetric since for instance
$[\Gamma_0 D_F, \Gamma_{6}]=0$ and $\{\Gamma_{11}, \Gamma_{6}\}=0$ and any eigenstate of $\Gamma_0\,D_F$ is mapped by $\Gamma_6$ in an eigenstate with the same
eigenvalue and opposite Weyl chirality.
Thus we may relax the Weyl  condition writing
\be
\mathscr{D}_f = {\det}^{1/4}\, (\Gamma_0\,D_F),
\ee
where the determinant is defined
on real 32 components spinors.  Since  $\det\Gamma_7=1$  we have
\be
 {\det}\, (\Gamma_7\Gamma_0\,D_F) = {\det}\,(-s_1\,\Gamma_7\,\partial_0+\Gamma_0\,\partial_1-u\,\Gamma_8\,\Gamma_{1234}).
\ee
Denoting by $s_2$ the eigenvalue of $\Gamma_{1234}$
\be
\Gamma_{1234}\,\vartheta = s_2\,\vartheta,\qquad s_2\in\{-1,1\},
\ee
we can write
\be
\mathscr{D}_f = \prod_{s_2=\pm 1}\,{\det}^{1/4}\,(-s_1\,\Gamma_7\,\partial_0+\Gamma_0\,\partial_1-s_2\,u\,\Gamma_8)
 = \prod_{s_2=\pm 1}\,{\det}^{1/8}\,(\partial_0^2-\partial_1^2+u^2-s_2\,u'\,\Gamma_{08}).
\ee
Taking $\Gamma_{08} = \sigma_3\otimes \mathbb{I}_{8}$ which is
 possible on the space of definite $\Gamma_{1234}$ chirality, we find
\be
\mathscr{D}_f = \prod_{s=\pm 1}\,{\det}^2\,(\partial_0^2-\partial_1^2+u^2+s\,u').
\ee
The small  string  or small $q$  expansion of the potential $u$
($\kappa^2 = q+\frac{q^2}{2}+\cdots$, see section 6)
\ba
u &=& \sin \sigma\  {q}^{1/2}+(\frac{\sin \sigma}{2}-\frac{\sin ^3\sigma}{4}
) q^{3/2}+(\frac{\sin ^5\sigma}{16}-\frac{17 \sin ^3\sigma
   }{64}+\frac{11 \sin \sigma}{32}) q^{5/2}+\cdots, \nonumber \\
u' &=& \cos \sigma\  {q}^{1/2}+(\frac{3 \cos ^3\sigma}{4}-\frac{\cos \sigma}{4}) q^{3/2}+(\frac{5 \cos ^5\sigma}{16}+\frac{11 \cos ^3\sigma
   }{64}-\frac{9 \cos \sigma}{64}) q^{5/2}+\cdots, \nonumber \\
u^2 &=& \sin ^2\sigma\   q+(\sin ^2\sigma-\frac{\sin ^4\sigma}{2}) q^2+(\frac{3 \sin ^6\sigma}{16}-\frac{25 \sin ^4\sigma}{32}+\frac{15 \sin
   ^2\sigma}{16}) q^3+\cdots \ .
\ea
Let  mention also that the  bosonic  fluctuation masses in conformal gauge discussed in section 6 have
the following expansions in the one-spin case  when $w_1=0, \ Q_1=0$
($\kappa^2 = q+\frac{q^2}{2}+\cdots$, see section 6)
\ba
M_X^2    = \left(2 \sin ^2\sigma-1\right) q+(-\sin ^4\sigma+2 \sin ^2\sigma-\frac{1}{2}) q^2
+(\frac{3 \sin ^6\sigma}{8}-\frac{25 \sin ^4\sigma
   }{16}+\frac{15 \sin ^2\sigma}{8}-\frac{11}{32}) q^3+... \no  \\
M_\eta^2 =-1+(2 \sin ^2\sigma-\frac{1}{2}) q+(-\sin ^4\sigma+2 \sin ^2\sigma-\frac{11}{32}) q^2
 + (\frac{3 \sin ^6\sigma}{8}-\frac{25
   \sin ^4\sigma}{16}+\frac{15 \sin ^2\sigma}{8}-\frac{17}{64}) q^3+... \no  \\
M_1^2    =\left(2 \sin ^2\sigma-1\right) q+(-\sin ^4\sigma+2 \sin ^2\sigma-\frac{1}{2}) q^2
+ (\frac{3 \sin ^6\sigma}{8}-\frac{25 \sin ^4\sigma
   }{16}+\frac{15 \sin ^2\sigma}{8}-\frac{11}{32}) q^3+\cdots, \no \\
M_2^2    =-1+(2 \sin ^2\sigma-\frac{3}{2}) q+(-\sin ^4\sigma+2 \sin ^2\sigma-\frac{27}{32}) q^2
+ (\frac{3 \sin ^6\sigma}{8}-\frac{25
   \sin ^4\sigma}{16}+\frac{15 \sin ^2\sigma}{8}-\frac{39}{64}) q^3+... \no  \\
Q_2=-2+(\sin ^2\sigma-\frac{1}{2}) q+(-\frac{1}{4} \sin ^4\sigma+\frac{3 \sin ^2\sigma}{4}-\frac{9}{32}) q^2
+ (\frac{\sin ^6\sigma
   }{16}-\frac{11 \sin ^4\sigma}{32}+\frac{39 \sin ^2\sigma}{64}-\frac{25}{128}) q^3+... \no \ea

\renewcommand{\theequation}{D.\arabic{equation}}
\renewcommand{\thesection}{D}
 \setcounter{equation}{0}
\setcounter{section}{1} \setcounter{subsection}{0}
\section{Perturbative computation of   stability angles    for pulsating string in $AdS_{3}$}
\label{appD}

Let us start with the bosonic
 type I fluctuations.
Setting in Eq.~(\ref{PulsAdSFluctI})
\be
\tau = \frac{2\mathbb K(\frac{R_{+}}{R_{+}-R_{-}})}{\pi}\,\frac{1}{\sqrt{R_{+}-R_{-}}} \,y,\qquad 0\le y\le 2\pi,
\ee
and expanding in $\k=\E_0\to 0$, we obtain
\be
\mathcal O_{I} = \mathcal O_{I,0}+\mathcal O_{I,1}\k^{2} + \dots,
\ \ \ \ \ \
\mathcal O_{I,0} = -\partial_{y}^{2}-n^{2},\  \quad
\mathcal O_{I,1} = -\frac{3}{2}\partial_{y}^{2}-z\sin^{2}y, \quad \dots.
\ee
The evaluation of the stability angle for these operators is very simple and leads to the results
in Eqs.~(\ref{PulsAdSStabI}).

For the bosonic type II fluctuations
 we get
\be
\mathcal O_{II} = \mathcal O_{II, 0}+\mathcal O_{II, 1}\k^{2} + \dots,
\ \ \ \ \ \
\mathcal O_{II, 0} = -\partial_{y}^{2}-n^{2}+\frac{2}{\sin^{2}y}, \ \ \ \ \ \
\mathcal O_{II, 1} = -\frac{3}{2}\partial_{y}^{2}-2+\cos 2y+\frac{3}{\sin^{2}y}.
\ea
The fluctuation equation for the $\zeta$ field at  leading order in $\k\to 0$ limit is (for $m=1$)
\be
\Big[-\partial_{ y}^{2}-n^{2}+\frac{2}{\sin^{2} y}\Big]\zeta^{(0)}_{n} = 0.
\ee
We can look for a periodic solution such that $\zeta^{(0)}_{n}\sin y\sim \widetilde \rho$ is smooth. One finds
 one solution for $n=0,1$ and two solutions for $n\ge 2$:
\ba
\zeta^{(0)}_{0} \sim \cot y, \ \ \ \ \
\zeta^{(0)}_{1} \sim  \csc y, \ \ \ \ \ \
\zeta^{(0)\pm}_{n\ge 2} \sim \sqrt{\sin y} \,{P}^{\pm\frac{3}{2}}_{n-\frac{1}{2}}(\cos y),
\ea
where $P_{n}^{m}$ are associated Legendre polynomials. The first few cases for $n\ge 2$ are
\ba
&&\zeta^{(0)+}_{2} \sim \csc y\,(\cos 3 y-3\cos y), \ \ \  \ \ \ \
\zeta^{(0)-}_{2} \sim\sin^{2} y,\ \ \ \  \\
&&\zeta^{(0)+}_{3} \sim \csc y\,(\cos 4 y-2\cos2 y), \ \ \ \ \ \
\zeta^{(0)-}_{3} \sim\sin^{2} y\cos y, \\
&&
\zeta^{(0)+}_{4} \sim \csc y\,(3\cos 5 y-5\cos3 y), \ \ \ \ \ \ \ \
\zeta^{(0)-}_{4} \sim\sin^{2} y(2+3\cos2 y) .
\ea	
The general solution for $\zeta^{(0)+}$ can be shown to be
\ba
\zeta^{(0)+}_{n} &\sim& \csc y\,\Big[\cos (n+1) y+\frac{n+1}{1-n}\cos(n-1) y\Big],
\ea
while the  general solution for $\zeta^{(0)-}$ is less explicit,
\ba
\zeta^{(0)-}_{n} &\sim& \mathop{\sum_{0\le p\le n}}_{p-n\in 2\mathbb Z}c_{p}\cos p y,
\ea
with certain coefficients $c_{p}$.
The idea is now to do perturbation theory in $\k$ hoping to find
closely related stability angles for $\zeta^{(0)\pm}_{n}$.
We can  consider a  perturbative expansion starting with the  linear combination
$
\zeta_{n}^{(0)+}+\mu_{n}\zeta^{(0)-}_{n}
$,
\be
\zeta_{n} = \zeta_{n}^{(0)+}+\mu_{n}\zeta^{(0)-}_{n} + \k^{2} \zeta_{n}^{(1)}+\cdots\quad .
\ee
The mixing coefficient $\mu_{n}$ is determined by the requirement that $\zeta_{n}$ is quasiperiodic at order $\k^{2}$.
In general, we  find
\be
\frac{\zeta_{n}(a+2\pi)}{\zeta_{n}(a)} = 1+i\,\k^{2}\nu_{n}^{(1)}+...   \   .
\ee
We cannot compute $\nu_{n}^{(1)}$ in a closed form as a function of $n$ because we do not have an explicit
expression for $\zeta_{n}^{(0)-}$ as a closed function of $n$. Nevertheless, we can work out the procedure
for several $n$ and try a simple rational function of $n$. This works very well and the result for the
expansion of the stability angle
\be\la{e4}
\log\frac{\zeta_{n}(a+2\pi)}{\zeta_{n}(a)} = i\,\k^{2}\nu_{n}^{(1)}+i\,\k^{4}\nu_{n}^{(2)}+
i\,\k^{6}\nu_{n}^{(3)}+...
\ee
agrees with the expressions in (\ref{PulsAdSStabII})

Let us mention  that for  the expansion of
 the fermionic operator  $\mathcal{O}_{III}$  from section 5.2  we get
\ba
&&\mathcal O = \mathcal O_{0}+\mathcal O_{1/2}\mathcal E + \mathcal O_{1}\mathcal E^{2} + \dots,\\
&&
\mathcal O_{0} = -\partial_{y}^{2}-n^{2}, \qquad
\mathcal O_{1/2} = \pm \, i\cos y, \qquad
\mathcal O_{1} = -\frac{3}{2}\partial_{y}^{2}-\sin^{2}y, \quad\dots
\ea

\renewcommand{\theequation}{E.\arabic{equation}}
\renewcommand{\thesection}{E}
 \setcounter{equation}{0}
\setcounter{section}{1} \setcounter{subsection}{0}
\section{On the  expression for one-loop energy in terms of stability angles}
\label{appE}

Here  we discuss at a level heuristic how one may obtain  the semiclassical result (\ref{pou})
from one-loop effective action in the  path integral approach.

Let us first consider  the case  of a stationary 2d
soliton for which the fluctuation Lagrangian may have only $\s$-dependent coefficients.
Then the  1-loop correction to the 2d energy  can be found  by
computing the 1-loop {\it Euclidean}  partition function\foot{We consider the theory on a Euclidean cylinder $R_{\tau_e} \times S^1_\sigma$, \ $\tau_e= i \tau$.
}
\be\la{pqt}
E_{2d} = \frac{1}{2\T_\infty}\log\det\big[-\partial_{\tau_e}^{2}-\partial_{\sigma}^{2}+V(\sigma)\big] \ ,
\ee
where $\T_\infty\to \infty $ is an arbitrary time interval.
 Taking the trace over  functions $\sim e^{i\omega\tau_{e}}$, we get
\be\la{yyyy}
E_{2d}= \frac{1}{2\T_\infty}\times \T_\infty \times \int_{\mathbb R}\frac{d\omega}
{2\pi}\log\det\big[\omega^{2}-\partial_{\sigma}^{2}+V(\sigma)\big]
=\frac{1}{2} \int_{\mathbb R}\frac{d\omega}{2\pi}\sum_{n}\log\big(\omega^{2}+\omega_{n}^{2}\big) \ ,
\ee
where the characteristic frequencies $\omega_{n}$ are  the eigenvalues of
$-\partial_{\sigma}^{2}+V(\sigma)$.
In general, one has ($R \to \infty$)
\be
\int_{-R}^{R} {d\omega}\ \log(\omega^{2}+\omega_{n}^{2}) =
-4R (1-\log  R) + 2 \pi  \omega_{n}+ \mathcal O(R^{-1}) \ .
\ee
Summing over bosons and fermions and ignoring the divergent terms (that will cancel in the present superstring case)
we then get the familiar expression
\be
E_{2d}
=\frac{1}{2}\sum_{n}(-1)^{F}\omega_{n} \ . \la{stat}
\ee
Let us  also  review a
different representation for the
determinant of the 1d operator like  $-\partial_{\sigma}^{2}+V(\sigma)$
with periodic boundary conditions (see, e.g., section 4 of \ci{bd}  for a summary).
Using general notations, consider the problem
\begin{equation}\label{rfer}
 \big[-\del_x^2 +V(x) \big]\,f(x) =\Lambda\, f(x) \ , \ \ \ \ \ \ \
 V(x+L)=V(x)  \ .
\end{equation}
Its two independent solutions  $f_\pm (x)=e^{\pm i \, p(\Lambda) \, x}\,
\chi_\pm (x), \ \  \chi_\pm (x+L)= \chi_\pm (x )$   satisfying
 \begin{eqnarray}
 f_\pm(x+L)=e^{\pm i\nu }\ f_\pm(x) \ , \ \ \ \ \ \  \nu = p L  \ ,
 \label{ftum}
 \end{eqnarray}
define  $p$, the  ``quasi-momentum'', and we also call $\nu$ the  ``stability angle''. In general,  $p$  is a function of $\Lambda, L$ and
a functional of $V$.  Then the determinant
of the above operator computed with periodic boundary conditions on the  eigen-functions
can be represented as\foot{Here we included normalization to the  free  operator determinant that we will ignore in what follows (the corresponding
constant factor  will cancel in a superstring combination of determinants).}
\be\label{tum}
\log   \det [-\del_x^2+V(x)-\Lambda ]- \log \det [-\del_x^2]  = \ln [-4 \sin^2 ( \frac{1}{2} \nu)]   \ . \ee

Let us now turn to the case of interest in this current paper:
 a  time-dependent solution, periodic in real time. The idea  is
again to rotate  to Euclidean time and take the  infinite time interval  limit
and  interpret
the energy as a ``ground-state'' energy in the path integral context.
Let us start   with the logarithm of the  real-time  partition function on time interval $r \T$\  ($r \to \infty$)
where $\T$ is
the period,
 \be  \frac{1}{2}     \log\det\big[\partial_{\tau}^{2}-\partial_{\sigma}^{2} +  U(\tau) \big]   =
   \frac{1}{2} \sum_{n=-\infty}^\infty \log\det \big[ \partial_{\tau}^{2}+n^{2}  + U(\tau)\big] \ .  \la{gh}
\ee
Here the sign of the potential $U$ is chosen so that it is positive in the free massive particle case. We may then
 reduce  the 2d determinant to a 1d one
by  using the  Fourier transform  ($\partial_{\sigma} \to in$).
Then $E_{2d}$   may be defined  by the  Euclidean rotation  of the   above expression \rf{gh}
divided by  $r \T$. We may then
use  the  above representation \rf{tum} for  the 1d determinant  in terms of the  stability angle $\nu$.
Noting that
(i)  going to Euclidean time suggests  to set $\nu \to i \nu$   and
(ii) since we are on the  interval $r \T$
 the accumulated stability angle will   get  a factor of $r$,
 we then  finish with  ($\nu \to i r  \nu$)
\be
E_{2d}  = \lim_{r\to \infty}\frac{1}{2r \T}
\sum^\infty_{n=-\infty}\log\big[4\sinh^{2}\frac{r \nu (n)}{2}\big]
 = \frac{1}{2\T}    \sum^\infty_{n=-\infty}       \nu (n)\  ,
\ee
where $\nu$ is stability angle of the real-time problem on the period $\T$ of the potential.
This heuristic derivation reproduces   the expression in \rf{pou}.


\renewcommand{\theequation}{F.\arabic{equation}}
\renewcommand{\thesection}{F}
 \setcounter{equation}{0}
\setcounter{section}{1} \setcounter{subsection}{0}

\section{Comments on    periodicity condition for fermions}
\label{appF}

As discussed  at the end of section 7, in \ci{mikh} it was suggested
that for pulsating strings
with odd  winding number the  fermions
(defined using   angular coordinates as tangent-space directions)
should  be chosen to be antiperiodic in $\s$.
Below  we shall   comment on  possible reason for
that  from flat space perspective  and then
present arguments   against  this  interpretation in our curved space case
by considering more general  case with non-zero orbital momentum in $S^5$.




\subsection{Pulsating string solution in flat space}

Let us start with pulsating solution in flat space.
In cartesian coordinates
\begin{equation}
ds^2 = -dt^2 + d x^2 + d y^2
\end{equation}
the pulsating solution is (this is of course
 the flat space limit of the $S^2$ pulsating solution of section 2.1)
\begin{equation}
t =\kappa \tau, \qquad \quad x = \frac{\kappa}{m} \sin m \tau \cos m \sigma,
\qquad \quad y= \frac{\kappa}{m}\sin m \tau \sin m \sigma
\end{equation}
The $ \theta^1 = \theta^2$ \ $\kappa$-gauge fixed quadratic fermionic term in the
GS action
 in flat space in cartesian coordinates is then  (for simplicity in this section we define the fermionic Lagrangian without the overall factor of $i$)
 \be
&&L= 2 \bar{\theta} D_F \theta, \qquad \qquad D_F= - \rho_0 \partial_0 + \rho_1 \partial_1
\ , \\
&&\rho_0= \kappa \Gamma_0 +\kappa \cos m \tau \cos m \sigma \Gamma_7 + \kappa \cos m \tau \sin m \sigma \Gamma_9, \ \
\rho_1= - \kappa \sin m \tau \sin m \sigma \Gamma_7 + \kappa \sin m \tau \cos m \sigma \Gamma_9
\nonumber
\ee
where we  labelled  the coordinate $x$ as $7$ and $y$ as $9$.
We can get rid of the $\sigma$ dependence in $D_F$ by using the rotation
\begin{equation}
\theta = e^{-\frac{m \sigma}{2} \Gamma_7 \Gamma_9} \tilde{\theta} \la{jis}
\end{equation}
leading to
\begin{equation}
 \tilde{D}_F = - ( \Gamma_0 + \cos m \tau  \Gamma_7)\partial_0 +  \Gamma_9 \sin m \tau  \partial_1+ \frac{m}{2}\sin m \tau \Gamma_7
\la{kq}\end{equation}
One can then put the
fermionic Lagrangian in the standard  free massless fermion
form
\begin{equation}
\bar D_F= - \Gamma_0 \partial_0 +\Gamma_9 \partial_1 \la{stas}
\end{equation}
using the redefinition (local boost and rescaling)
\begin{equation}
\tilde{\theta} = \sqrt{\cosh q}\
 e^{-\frac{1}{2}q \Gamma_0 \Gamma_7}
 \bar{\theta},\quad  \qquad \cosh q = \frac{1}{|\sin m
 \tau|}
\la{qqk}\end{equation}
Note that the rotation \rf{jis}
changes periodicity of the fermions: if we start with periodic $\theta$
we get $\tilde \theta$ (and thus also $\bar \theta$)  antiperiodic
for odd $m$.

Let  us now repeat the same computation   starting with
the same pulsating solution written  in polar coordinates,
\begin{equation}\la{qqs}
ds^2 = -dt^2 + d \psi^2 + \psi^2 d \phi^2
\ , \ \ \ \
t =\kappa \tau, \quad \quad \psi= \frac{\kappa}{m} \sin m\tau, \quad \quad \phi=m \sigma
\end{equation}
This is again  the short string limit of the pulsating solution in $S^2$
in \rf{soll}. The bosonic part of the fluctuation  Lagrangian
is trivial, while the quadratic part of the
GS superstring  action written in general coordinates
\begin{equation}
L= (\sqrt{-g}g^{ab}\delta^{IJ}- \epsilon^{ab}s^{IJ})\bar{\theta}^{I} \rho_a D_b \theta^{J} \ .
\la{llc}
\end{equation}
takes  the following form in the  $\theta^1=\theta^2$   gauge
(here we use labels $0$ for $t$, $7$ for $\psi$, and $9$ for $\phi$)\foot{One finds
$
\rho_0 = \kappa \Gamma_0 +\kappa \cos m \tau \Gamma_7 ,
\quad  \rho_1= \kappa \sin m \tau \Gamma_9
$
and $
D_0 = \partial_0,  \quad D_1=\partial_1- \frac{m}{2}\Gamma_{79}
$.}
\bea
&&L= 2 \bar{\theta} D_F \theta \ , \ \ \ \ \ \ \ \ \ \ \ \ D_F= - \rho_0 D_0 + \rho_1 D_1 \ ,
\\
&& D_F= - ( \Gamma_0 +  \cos m \tau  \Gamma_7)\partial_0 +
\Gamma_9 \sin m \tau  \partial_1+ \frac{m}{2} \sin m \tau \Gamma_7
\eea
This operator is the same as in  \rf{kq}, so to put it in the standard form \rf{stas}
one needs  again the same  local boost and rescaling as in \rf{qqk}.
Since \rf{qqk}  does not change periodicity of the fermions,
that seems to imply that to match the cartesian coordinate choice result,
starting with GS action in coordinates \rf{qqs}
we need to assume that fermions are antiperiodic for odd $m$.

That conclusion  may seem strange as
 we need standard periodic fermions to cancel  corrections to ground-state energy.
Also, it seems strange to assume that the choice of   periodic/antiperiodic
boundary conditions for the  fermions in the original action \rf{llc}
may depend on a specific choice of the bosonic solution:
for any  bosonic background the flat-space
GS fermions are free (and periodic on a cylinder)
 in the  light-cone gauge \ci{grs}, so that
 periodicity/antiperiodicity  issue  is likely to   be  a  gauge/coordinate  artifact.
Indeed,  starting
 with the  above solution in either cartesian or polar coordinates
and writing the  GS action \rf{llc} in the light-cone gauge
$
\Gamma_{+} \theta^I=0,  \quad \Gamma_{\pm}= \frac{1}{2}(\mp \Gamma_0 + \Gamma_7)
$
one  ends up  with the same free operator \rf{stas}  defined
for  either periodic or antiperiodic fermions  which appears
to contradict the above conclusions.

Let us discus explicitly what one finds in the  light-cone gauge.

\

\underline{Light-cone gauge in  polar coordinates}

In polar coordinates we have
\be  &&
\rho_0 = \kappa \Gamma_0 +\kappa \cos m \tau \Gamma_7 , \quad \quad \rho_1= \kappa \sin m \tau \Gamma_9\ ,
\nonumber \\
 &&D_0 = \partial_0, \quad \quad D_1=\partial_1- \frac{m}{2}\Gamma_{79}
\ee
We fix the $\kappa$ symmetry as
\begin{equation}\la{ccl}
\Gamma_{+} \theta^I=0, \quad \quad \Gamma_{\pm}= \frac{1}{2}(\mp \Gamma_0 + \Gamma_7)
\end{equation}
The the fermionic Lagrangian is then
\begin{equation}
L= - \bar{\theta}^1 (1+\cos m \tau) \Gamma_- (\partial_0+ \partial_1) \theta^1 - \bar{\theta}^2 (1+\cos m \tau) \Gamma_- (\partial_0- \partial_1) \theta^2 + \frac{m}{2}\bar{\theta}^I \sin m \tau \Gamma_- \theta^I
\end{equation}
Performing  the rescaling
\begin{equation}
\theta^I \rightarrow \frac{\tilde{\theta}^I}{\sqrt{1+\cos m \tau}}
\end{equation}
we obtain
\begin{equation}
L= - \bar{\theta}^1 \Gamma_- (\partial_0+ \partial_1) \theta^1 - \bar{\theta}^2 \Gamma_- (\partial_0- \partial_1) \theta^2
\end{equation}
Thus starting with periodic fermions  we get free action with periodic
fermions.

\

\underline{Light-cone gauge with cartesian coordinates}

In cartesian coordinates we have
\be
\rho_0 = \kappa \Gamma_0 + \kappa \cos m \tau \cos m \sigma \Gamma_7 + \kappa \cos m \tau \sin m \sigma \Gamma_9
\nonumber \\
\rho_1= -\kappa \sin m\tau \sin m \sigma \Gamma_7 + \kappa \sin m \tau \cos m \sigma \Gamma_9, \quad \quad D_0=\partial_0, \quad \quad D_1=\partial_1
\ee
Fixing again the same light-cone gauge \rf{ccl}
we get the  fermionic Lagrangian
\begin{eqnarray}
L_F=&-&\bar{\theta}^1 (1+ \cos m \tau \cos m \sigma + \sin m \tau \sin m\sigma) \Gamma_- (\partial_0+\partial_1)\theta^1 \nonumber\\
&-& \bar{\theta}^2 (1+ \cos m \tau \cos m \sigma - \sin m \tau \sin m \sigma) \Gamma_- (\partial_0 -\partial_1)\theta^2
\end{eqnarray}
Rescaling the fermions as
\begin{equation}
\theta^1 \rightarrow \frac{\tilde{\theta}^1}{\sqrt{1+\cos m \tau \cos m \sigma+ \sin m \tau \sin m \sigma}}, \quad \quad
\theta^2 \rightarrow \frac{\tilde{\theta}^2}{\sqrt{1+\cos m \tau \cos m \sigma - \sin m \tau \sin m \sigma}}
\end{equation}
we obtain
\begin{equation}
L= - \bar{\theta}^1 \Gamma_- (\partial_0+ \partial_1) \theta^1 - \bar{\theta}^2 \Gamma_- (\partial_0- \partial_1) \theta^2
\end{equation}
Once  again,  starting with periodic fermions
we end up with  periodic fermions. This is in contrast
to what happened in  the $\theta_1=\theta_2$ gauge where a $\sigma$-dependent rotation was needed, which changed the periodicity of fermions.

\subsection{Pulsating string solution   with extra angular momentum $J$ on $S^5$}

To clarify what is going on further let us consider
a generalization of the  $S^2$ pulsating solution  to the presence of angular momentum $J$
along a direction of $S^5$ transverse to $S^2$  \ci{jmz,krut}.
That will  allow us to interpolate to large values of $J$  and  thus
resolve the question about the fermionic boundary conditions
by comparing to the  BMN limit.

Starting with the metric of $\mathbb{R} \times S^3$
\begin{equation}
ds^2 = -dt^2 + d \psi^2 + \sin^2 \psi d \phi^2 + \cos^2 \psi d \varphi^2
\end{equation}
the solution with non-zero $J=\sql \mathcal{J}$ is
\be
&&t=\kappa \tau, \quad \quad \psi=\psi(\tau), \quad \quad \varphi=\varphi(\tau), \quad \quad
\phi=m \sigma\ , \\
&&\dot{\psi}^2 +m^2 \sin^2 \psi + \frac{\mathcal{J}^2}{\cos^2 \psi}=\kappa^2\ , \ \ \ \ \ \ \
\mathcal{J}=\cos^2 \psi \dot{\varphi}= {\rm const}
 \ . \ee
Then in the quadratic part of the \adss GS  Lagrangian
\be &&
L= i (\eta^{ab} \delta^{IJ} - \epsilon^{ab} s^{IJ}) \bar{\theta}^I \rho_a \mathcal{D}_a
\theta^J \ , \la{fet}\\
&&
\mathcal{D}_a \theta^I= D_a \theta^I - \frac{i}{2}\epsilon^{IJ} \Gamma_* \rho_a \theta^J,
 \quad \quad \quad s^{IJ}=(1,-1), \quad \quad \Gamma_*=i \Gamma_{01234} \la{ffet}
\ee
we have
\begin{equation}
\rho_0= \kappa \Gamma_0 + \dot{\psi}\Gamma_7 + \cos \psi \dot{\varphi}\Gamma_8, \quad \quad \rho_1=m \sin \psi \Gamma_9
\end{equation}
\begin{equation}
D_0= \partial_0 + \frac{1}{2}\sin \psi \dot{\varphi} \Gamma_{78}, \quad \quad D_1= \partial_1 - \frac{m}{2}\cos \psi \Gamma_{79}
\end{equation}
where we label the coordinates as $7$ for $\psi$, $8$ for $\varphi$ and $9$ for
$\phi$.
The resulting fermionic operator in the $\theta^1 = \theta^2$ gauge  is
\begin{eqnarray}
D_F&=& (\kappa \Gamma_0 + \dot{\psi}\Gamma_7 + \cos \psi \dot{\varphi}\Gamma_8) \partial_0- m \sin \psi \Gamma_9 \partial_1 + \frac{1}{2}\sin \psi \dot{\varphi} (\kappa \Gamma_{078} + \dot{\psi}\Gamma_8 -\cos \psi \dot{\varphi}\Gamma_7)\nonumber\\
& -& \frac{m^2}{2}\sin \psi \cos \psi \Gamma_7
\pm  m \sin \psi \Gamma_{09} (\dot{\psi}\Gamma_7 + \cos \psi \dot{\varphi}\Gamma_8)
\end{eqnarray}
where we projected onto  eigenspaces with  $\Gamma_{1234}=\pm 1$.
Performing two boosts  -- in the $(07)$ plane
as
\begin{equation}
\theta = e^{-\frac{1}{2}\alpha \Gamma_0 \Gamma_7} \tilde{\theta}, \qquad\qquad
\cosh \alpha = \frac{\kappa}{\sqrt{\kappa^2-\dot{\psi}^2}}
\end{equation}
and in the $(08)$ plane
\begin{equation}
\tilde{\theta} = e^{-\frac{1}{2}\beta \Gamma_0 \Gamma_8} \bar{\theta},
 \qquad\qquad  \cosh \beta = \frac{\sqrt{m^2 \sin^2 \psi + \cos^2 \psi \dot{\varphi}^2}}{m \sin \psi}
\end{equation}
and rescaling
\begin{equation}
\bar{\theta}=\frac{1}{\sqrt{|m \sin \psi|}}\hat \theta   \label{resc}
\end{equation}
we end up with $ D'_F  = s   \hat  D_F, \ \  s\equiv {\rm sign}(\sin \psi)  $,
where
\begin{eqnarray}
\hat D_F &=& \Gamma_0 \partial_0 - \Gamma_9 \partial_1 -  \frac{\kappa m}{2}\frac{\dot{\varphi} \cos(2 \psi)}{m^2 \sin^2 \psi+ \cos^2 \psi \dot{\varphi}^2} \Gamma_{078} \\
&\pm & \frac{m \dot{\psi} \sin \psi}{\sqrt{m^2 \sin^2 \psi + \cos^2 \psi
\dot{\varphi}^2}}\Gamma_{097} \pm \frac{\kappa \dot{\varphi}\cos \psi}{
\sqrt{m^2 \sin^2 \psi + \cos^2 \psi \dot{\varphi}^2}}\Gamma_{098}\nonumber
\end{eqnarray}
In the BMN limit of small string with large orbital momentum, i.e.
 $m\to 0, \ \psi \rightarrow 0$, $\kappa=\mathcal{J}$,  we get
the standard result
\begin{equation}
\hat D_F=\Gamma_0 \partial_0 - \Gamma_9 \partial_1 \pm \mathcal{J} \Gamma_{098} \ . \la{koll}
\end{equation}
In the limit of $\mathcal{J}\rightarrow 0$ we end up with
\begin{equation}
\hat D_F=  \Gamma_0 \partial_0 - \Gamma_9 \partial_1 \pm  s \dot{\psi} \Gamma_{097} \la{moll}
\end{equation}
This is essentially the same as one finds by starting directly with $J=0$
as in Appendix A.
Since the fermions must be periodic  to match the BMN limit \rf{koll},
they should also be periodic  in the opposite pulsating string limit \rf{moll}.
While formally  the transition between $m=0$ and $m\not=0$ cases may
still look discontinuous, on physical grounds  it seems natural to
expect that near-BMN state
represented  by small pulsating string with large $J$
should  belong to a family of solutions  that should all be quantized
with periodic fermions.

Let us now  consider the same  computation  choosing  the
 light-cone gauge
\begin{equation}\Gamma_+\theta^I=0\ , \ \ \ \ \
\Gamma_{\pm} \equiv \frac{1}{2}(\mp \Gamma_0 + \Gamma_8),
 \quad \quad \Gamma_+ \Gamma_- + \Gamma_- \Gamma_+=1
\end{equation}
in the  Lagrangian \rf{fet}.
Performing  the rescaling
\begin{equation}
\theta^I= \frac{\tilde{\theta}^I}{\sqrt{\kappa+\dot{\varphi}\cos \psi}}
\end{equation}
we  obtain (omitting tilde on $\theta$)
\begin{eqnarray}
L= &-& \bar{\theta}^1  \Gamma_- (\partial_0+\partial_1) \theta^1
- \bar{\theta}^2 \Gamma_- (\partial_0-\partial_1) \theta^2 + \frac{m}{2}\frac{\kappa \cos \psi + \dot{\varphi}\cos (2 \psi)}{\kappa+ \dot{\varphi} \cos \psi}s^{IJ} \bar{\theta}^I \Gamma_- \Gamma_{79} \theta^J \nonumber\\
&+& 2 \kappa \bar{\theta}^1 \Gamma_- \Pi \theta^2 - \frac{2 m^2 \sin^2 \psi}{\kappa+ \dot{\varphi}\cos \psi} \bar{\theta}^1 \Gamma_- \Pi \theta^2+
\frac{2 m \dot{\psi} \sin \psi}{\kappa+ \dot{\varphi} \cos \psi} \bar{\theta}^1 \Gamma_- \Gamma_{79} \Pi \theta^1
\end{eqnarray}
where the second line comes from the second  RR coupling term in
\rf{ffet}  and   $\Pi=\Gamma_{1234}$.
For  $
m \rightarrow 0,  \quad \psi \rightarrow 0,  \quad \kappa \rightarrow \mathcal{J}  $
we get
\begin{equation}
L= - \bar{\theta}^1  \Gamma_- (\partial_0+\partial_1) \theta^1
- \bar{\theta}^2 \Gamma_- (\partial_0-\partial_1) \theta^2+ 2 \mathcal{J} \bar{\theta}^1 \Gamma_- \Pi \theta^2
\end{equation}
which is the familiar
BMN expression. Note that for $\psi \rightarrow 0$ but
$m$ arbitrary we get an additional $m$-dependent term.
The limit $\mathcal{J} \rightarrow 0$ is smooth
and leads to
\begin{eqnarray}
L = &-& \bar{\theta}^1  \Gamma_- (\partial_0+\partial_1) \theta^1
- \bar{\theta}^2 \Gamma_- (\partial_0-\partial_1) \theta^2 + \frac{m}{2} \cos \psi s^{IJ} \bar{\theta}^I \Gamma_- \Gamma_{79} \theta^J
+ 2 \kappa \bar{\theta}^1 \Gamma_- \Pi \theta^2 \nonumber\\
&+&  \frac{2 m \dot{\psi} \sin \psi}{\kappa} \bar{\theta}^1 \Gamma_- \Gamma_{79} \Pi \theta^2
- \frac{2 m^2 \sin^2 \psi}{\kappa} \bar{\theta}^1 \Gamma_- \Pi \theta^2
\end{eqnarray}
To conclude,  embedding pulsating  solution into a
more general case with $J\not=0$ suggests that the fermion boundary conditions should be fixed
universally  rather than be sensitive to particular values of $m$.

\end{document}